\documentclass[10pt]{article}

\usepackage{newtxtext,newtxmath}
\usepackage{graphicx}

\usepackage[letterpaper,margin=1in]{geometry}

\usepackage{xr}
\renewenvironment{abstract}
	{\quotation}
	{\endquotation}

\date{}

\makeatletter
\renewcommand{\fnum@figure}{\textbf{Figure \thefigure}}
\renewcommand{\fnum@table}{\textbf{Table \thetable}}
\makeatother

\usepackage{scicite}

\usepackage{url}

\def\scititle{
	Hybridization of topologically distinct quartet modes in three-terminal graphene Josephson junctions
}
\title{\bfseries \boldmath \scititle}

\author{
	Asmaul~Smitha~Rashid$^{1,2}$,
	Le~Yi$^{3,2}$,
	Takashi~Taniguchi$^{4}$,
	Kenji~Watanabe$^{5}$,\and
	Nitin~Samarth$^{3,6,2}$,
	R\'egis~M\'elin$^{7}$,
	Morteza~Kayyalha$^{1,2\ast}$\and
	\small$^{1}$Department of Electrical Engineering, The Pennsylvania State University, University Park, Pennsylvania 16802, USA.\and
	\small$^{2}$Materials Research Institute, The Pennsylvania State University, University Park, Pennsylvania 16802, USA.\and
	\small$^{3}$Department of Physics, The Pennsylvania State University, University Park, Pennsylvania 16802, USA.\and
	\small$^{4}$Research Center for Materials Nanoarchitectonics, National Institute for Materials Science,  1-1 Namiki, Tsukuba 305-0044, Japan.\and
	\small$^{5}$Research Center for Electronic and Optical Materials, National Institute for Materials Science, 1-1 Namiki, Tsukuba 305-0044, Japan.\and
	\small$^{6}$Department of Materials Science and Engineering, The Pennsylvania State University, University Park, Pennsylvania 16802, USA.\and
	\small$^{7}$Université Grenoble-Alpes, CNRS, Grenoble INP, Institut NEEL, Grenoble, France.\and
	\small$^\ast$Corresponding author. Email: mzk463@psu.edu
}

\begin{document} 

% Insert the title and author list
\maketitle

% Abstract, in bold
% There are strict length limits, and not all formats have abstracts.
% Consult the journal instructions to authors for details.
% Do not cite any references in the abstract.
\begin{abstract} \bfseries \boldmath
Multiterminal Josephson junctions offer a powerful playground for exploring exotic superconducting and topological phenomena beyond the reach of conventional two-terminal devices. In this work, we present the direct spectroscopic observation of Cooper quartet resonances, a signature of correlated tunneling of two Cooper pairs across the device, in a graphene three-terminal Josephson junction (3TJJ).
Using tunneling spectroscopy, we visualize how Andreev bound states (ABS) evolve across a two-dimensional superconducting phase space, controlled by the two independent phase differences in the 3TJJ. These measurements reveal sharp local minima in the differential conductance spectra locked in a specific phase condition of superconducting phase variables. The resulting
  quantized trajectories around the compact torus of the
  superconducting phase variables reveal an underlying topological
  winding in the multipair transport. To interpret our results, we develop a theoretical model that connects the observed quartet resonances to the coherent hybridization of multiple ABS branches, a hallmark of the rich pairing process enabled by multiterminal geometries.
Our results highlight the potential of multiterminal superconducting devices to host engineered superconducting states and pave the way for new approaches to topological band structure design based on phase-controlled, higher-order superconducting transport.
\end{abstract}

\noindent
\section*{Introduction}
Josephson junctions (JJs) underpin a wide range of superconducting quantum technologies \cite{Furusaki1991, Kulik1970, Ishii1970, Bagwell1992,Kraft2018V2}, where coherent transport of Cooper pairs forms the basis for a multitude of device structures and basic physics studies, including Andreev level qubits \cite{Coraiola2023spin, Chtchelkatchev2003, Zazunov2003, Padurariu2010, HaysAndreevspin2021, PIta2023, Gingrich2016}, quantum sensors \cite{NazarovSpintronics, Lee2020}, and topological phases \cite{Quitopologicalreview}. These studies focus on JJs with two superconducting terminals, where the Andreev bound state (ABS) energies depend on a single superconducting phase difference \cite{Sauls2018, Nazarov2019}.

Multiterminal Josephson junctions (MTJJs) expand this landscape by adding independent superconducting phase differences that control  the energies of ABSs {\cite{Pankratova2020,Graziano2020, Draelos2019, Arnault2021, MatutequantumCircuit2023}}. These phase differences can be interpreted as quasi-momenta in a synthetic multi-dimensional crystal. The ABS spectra therefore resemble a band structure in higher dimensions, known as the Andreev band structure {\cite{ Riwar2016, Meyer2017,Nazarov2019,Peralta2019, Chandrasekhar2022, Peralta2023}}. These band structures host unique properties such as higher-order Cooper multiplets \cite{quartettomography2024}, Andreev molecules \cite{Kornichmolecule2019, Matsuophase2023, kocsis2024molecule, Pillet2019, Coraiola2023, Matsuophase2023, Coraiola2023spin, Arnault2021}, Floquet-ABS \cite{Oka2019, Liu2019, Park2022, Carrad2022}, nonequilibrium Andreev resonances \cite{NonequilibriumABS2025, Melin3TJJ} and topological phases \cite{{Strambini2016,Meyer2017,Nazarov2019,Peralta2019,Chandrasekhar2022,Riwar2016,Peralta2023,xie2018weyl}}, making MTJJs an unparalleled platform for studying exotic quantum states. 

Among these emergent processes, Cooper multiplets such as quartets, in which two Cooper pairs coherently split across three terminals, stand out as a hallmark signature of multiterminal entanglement \cite{quartettomography2024,  gupta2024evidence, Grazianoselective2022, Zhang2023, cohen2018nonlocal, huang2022evidence, Melin_2014, pfeffer2014subgap,Freyn2011}. The quartet processes give rise to supercurrents that depend on novel combinations of phase variables (e.g, $\chi_q$ = $2\varphi_L-\varphi_R-\varphi_B$
or $\chi_q=\varphi_L-2\varphi_R+\varphi_B$), where $\varphi_i$ ($i = L, B, R$) denote the superconducting phases of the terminals
$S_L$, $S_B$, and $S_R$, respectively and $\chi_q$ is the corresponding quartet phase variable. 
(Figure~\ref{Ch6_bisquuid.jpg}(a)). Theoretical work has proposed strategies to identify these multiplet states through current-voltage characteristics or supercurrent-phase measurements
  \cite{quartettomography2024,melin2016gate,rech2014}. Experimental
evidence \cite{gupta2024evidence, cohen2018nonlocal,
  Grazianoselective2022, Zhang2023}, though suggestive, remains
largely indirect and often relies on integrated transport signatures
that obscure the fine structure of the underlying ABS
spectrum. Therefore, these approaches lack the resolution to
resolve the structure of the underlying ABS
associated with higher-order Cooper processes. A spectroscopic
approach capable of isolating the ABS dispersions
over multi-dimensional phase space is essential for unambiguous
detection of 
  Cooper quartets. Recent developments in tunneling
  spectroscopy of MTJJs \cite{Pillet2019, Coraiola2023,
    Matsuophase2023, MatsuoScience2023, Coraiola2023spin,
    Carrad2022,Lee2025} offer a powerful route to directly probe
  these multi-dimensional Andreev bands, providing insight into the
  details of their spectral features and topological properties
  \cite{Lee2025,Tunnel4TJJ,Coraiola2023}. Yet, despite their
  potential, these studies have not been fully leveraged to explore
  the interplay of higher-order Cooper processes within the ABS
  spectrum.
  
  In this work, we utilized a combined experimental and theoretical approach to probe the Andreev band structure and report quantized winding
  trajectories that indicate nontrivial topology in multipair
  processes. We determine how these winding sectors manifest and hybridize in the
  ABS spectrum. Our findings address whether and how these multipair processes are related to the emerging topological viewpoint
  in MTJJs. We specifically utilize superconducting
tunneling spectroscopy in a three-terminal graphene JJ, accessing the
full Andreev band structure as a function of the two
independent superconducting phase differences. By fixing the tunnel
probe bias and sweeping the two phases, we resolve the
phase-dependent dispersion of ABSs.  We observe
  spectroscopic signatures associated with the Cooper
  quartets, specifically, sharp resonances along specific
  phase combinations and quantized slope in
  the plane of the superconducting phase variables.
  Furthermore, we find avoided crossings for two quartet modes
  indicative of ABS hybridization. This hybridization reveals the
  quantum mechanical nature of the quartets and shows
  that the topological winding structure alone is insufficient to
  describe the behavior of the Andreev bands. We compare our
measurements to theoretical calculations that incorporate coherent
tunneling processes in the MTJJ. The simulations qualitatively
reproduce the key spectral features including quartet
resonances supporting the overall quartet picture.
Our results thus provide direct, phase-resolved spectroscopic evidence
of higher-order Cooper quartet processes and establish graphene-based
MTJJs as a versatile platform for exploring exotic quantum states in
synthetic superconducting lattices.

\section*{Results}

\begin{figure}[ht]
\centering
\includegraphics[width=12 cm]{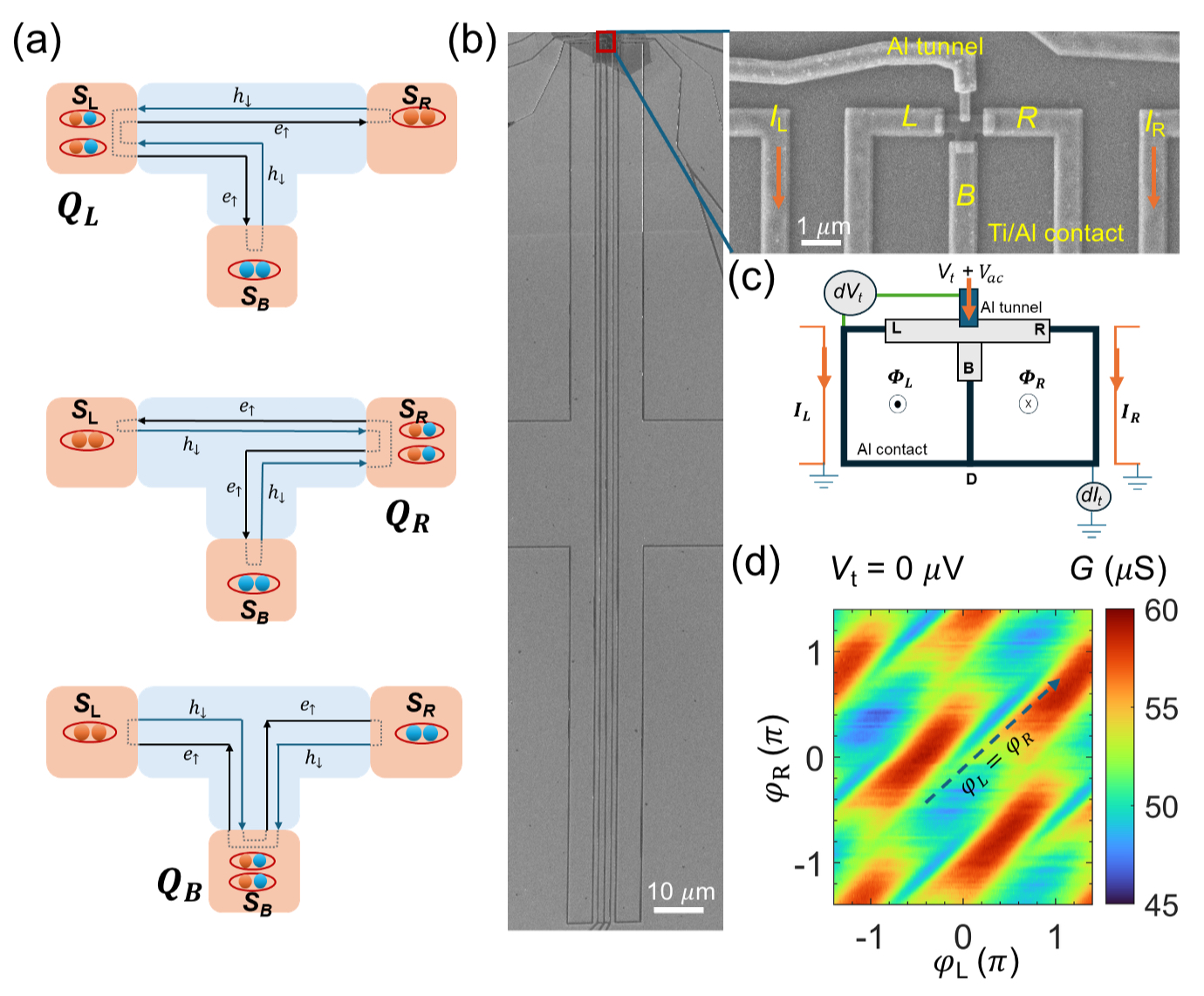}
 \caption{\label{Ch6_bisquuid.jpg} \textbf{Schematics of Cooper quartet processes and experimental device.} (a) $Q_L$, $Q_R$, and $Q_B$ schematics represent three-terminal quartet processes where electron-hole conversion occurs twice at either $S_L$, $S_R$, or $S_B$, respectively. $Q_L$, $Q_R$, and $Q_B$ have phase relations as $2\varphi_L-\varphi_R-\varphi_B$, $-\varphi_L+2\varphi_R-\varphi_B$
  and $-\varphi_L-\varphi_R+2\varphi_B$, respectively. (b) Scanning electron microscope (SEM) image of a representative three-terminal JJ. The magnified SEM image on the right highlights the three superconducting terminals denoted as $S_i$ ($i =  L, B, R$). A fourth Al terminal is weakly coupled to the device forming a tunnel probe $S_T$.  (c) Schematic representation of the device and the measurement setup. (d) Tunneling conductance $G$ as a function of phases $\varphi_L$ and $\varphi_R$ at bias voltage $V_t$ = 0 V. $\varphi_L$ and $\varphi_R$ are obtained by applying $I_L$ and $I_R$ to the flux bias lines with step sizes of 0.25 $\mu$A and 1 $\mu$A, respectively. }
\end{figure}

We fabricate three-terminal Josephson junctions on hBN/graphene/hBN van der Waals heterostructures, with edge-contacted superconducting terminals made of Ti (10~nm)/Al (80~nm). These terminals are coupled through the shared graphene region, allowing Josephson supercurrent to flow between them. The graphene channel is 0.6~$\mu$m wide, and each terminal is positioned approximately 0.6~$\mu$m from the center. Figure~\ref{Ch6_bisquuid.jpg}(b) shows a scanning electron microscope (SEM) image of a representative three-terminal junction. The three superconducting electrodes are labeled $S_i$ ($i = L, B, R$). These terminals are connected to a common node $D$, forming two closed superconducting loops.
Each terminal is characterized by its superconducting phase $\varphi_i$ ($i = L, B, R$). The Josephson current in the 3TJJ is governed by Andreev bands, whose energies depend on the phase differences between the terminals. Assuming $\varphi_B = 0$, two independent phase differences, $\varphi_{L/R} = 2\pi\Phi_{L/R}/\Phi_0$, controlled by external magnetic fluxes $\Phi_{L/R}$, determine the behavior of the junction. These fluxes are applied via two Ti (10 nm)/Al (80 nm) bias lines for the left and right loops. 
Finally, a superconducting tunneling probe $S_T$ is fabricated by edge-contacting graphene with Al (80 nm), without a Ti layer~\cite{Bretheau2017, mccumber1968tunneling, Gaviour1960, Lee2025}. This probe enables tomography of the Andreev bands in the graphene-based 3TJJ.

%\section{Two-dimensional Andreev bands}
 
Figure~\ref{Ch6_bisquuid.jpg} (c) shows the measurement setup for Device 1, which is shown in Fig.~\ref{Ch6_bisquuid.jpg} (b). 
The differential tunneling conductance is obtained as $G = {dI_t}/{dV_t}$ where the tunneling current (voltage) is measured between $S_T$ and node $D$. In this device, if the voltage $V_t$ on the tunneling
tip is smaller than the superconducting gap $\Delta$, then $G$ is
expected to have peaks that emerge in the tunneling current when the BCS gap-edge singularity in the density of states of the probe $S_t$ matches the ABS energies
\cite{mccumber1968tunneling, Gaviour1960,Falci_2001}. However,
we still obtain a nontrivial signal if the bias
voltage energy $eV_t$ is larger than twice the
  gap. In this case, quasiparticles exchange between graphene and $S_t$ will form a nonequilibrium Fermi surface at the chemical potential $\mu_N(V_t)$ 
\cite{Yip1998,Giantconductance1995,melin2025revisitingadiabaticlimitballistic}. The tunnel probe in Device~1 exhibits a soft superconducting gap (see section I in the Supporting Information (SI) for more details). For tunneling voltages within the gap ($|eV_t| < \Delta$, with $\Delta \approx 180\,\mu\text{eV}$), tunneling conductance directly measures the sub-gap Andreev bound states (ABSs). For voltages outside the gap ($|eV_t| > 2\Delta$), 
the probe accesses nonequilibrium ABSs via quasiparticle injection and relaxation in graphene, as described above.

\begin{figure}[ht]
\centering
\includegraphics[width=10 cm]{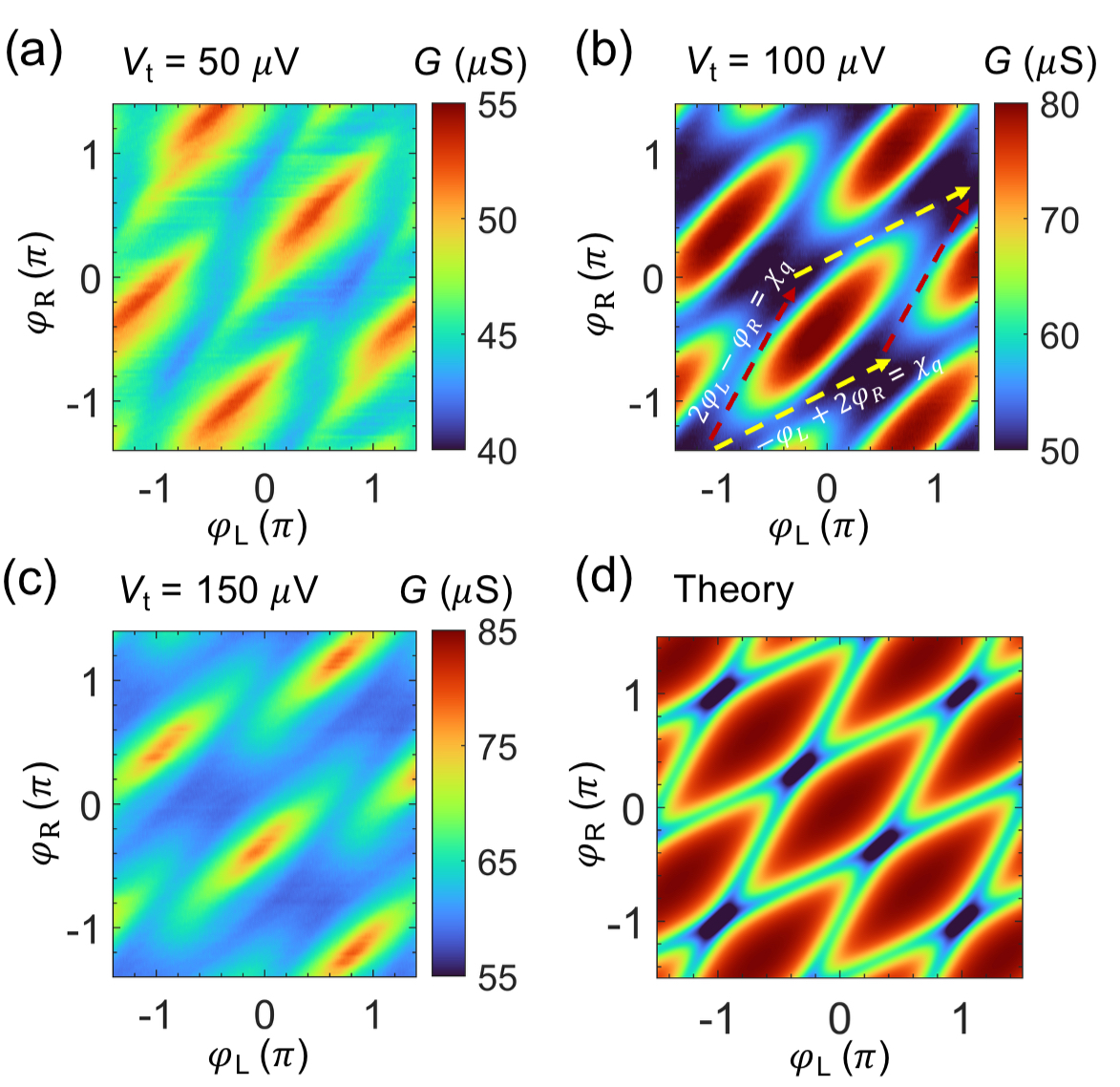}
 \caption{\label{Ch6_DOS.jpg} \textbf{Tunneling spectroscopy in
     three-terminal JJ.} (a-c) Tunneling conductance $G$ as a function
   of independent phases $\varphi_L$ and $\varphi_R$ governed by
   fluxes $\Phi_L$ and $\Phi_R$ at various bias voltages $ V_t$ = 50
   $\mu$V (a), $ V_t$ = 100 $\mu$V (b), and $ V_t$ = 150 $\mu$V (c). The
   colormaps are obtained by varying $I_L$ and $I_R$ with a step sized
   of 0.25 $\mu$A and 1 $\mu$A respectively. (d) Theoretically
   calculated tunneling conductance as a function of phases
   $\varphi_L$ and $\varphi_R$ at dimensionless probe
     bias voltage $V_{t,eff} = |eV_t|/2\Delta$.} 
\end{figure}

Figure~\ref{Ch6_bisquuid.jpg} (d) shows the tunneling conductance $G$ as a function of independent phases $\varphi_L$ and $\varphi_R$, tuned by fluxes $\Phi_L$ and $\Phi_R$, at $ V_t$ = 0 V. We observe periodic modulation of $G$ over $\varphi_L$ and $\varphi_R$. 
The minima in the $G$ map (referred to as resonances thereafter) follow a titled plane along $\varphi_L = \varphi_R$ direction. These resonances are attributed to the presence of ABS between ($S_L$, $S_B$) and ($S_R$, $S_B$) whose energies disperse with the superconducting phases. 
Figures~\ref{Ch6_DOS.jpg} (a-c) show the tunneling conductance $G$ as a function of $\varphi_L$ and $\varphi_R$ at constant $V_t$’s. For $V_t < \Delta$, $G$ exhibits periodic resonances corresponding to the presence of ABS between ($S_L$, $S_B$) and ($S_R$, $S_B$). However, as $V_t$ approaches $\Delta$, the lines corresponding to the resonances [dominant minima in Figs.~\ref{Ch6_DOS.jpg} (b-c)] realign along $\chi_q$ = $2\varphi_L-\varphi_R$  and $\chi_q$ = -$\varphi_L+2\varphi_R$. Furthermore, these two families of resonant minima (dark blue lines) create a diamond-like closed loop in $\varphi_L$, $\varphi_R$ phase space as shown by the dashed lines in Fig.~\ref{Ch6_DOS.jpg} (b). A recent study attributes such tilted resonant lines to an effect of the cross-talk between the left/right loops and the local magnetic field produced by the flux lines \cite{Coraiola2023}. However, we observe resonances of quantized
  slope $2$ or $1/2$, controlled by the bias voltage. This bias-voltage evolution of the tilt cannot be explained by the crosstalk. As we discuss theoretically below, these observations point to a microscopic origin tied to higher-order Andreev processes.

%%%%%%%%%%%%%

To understand the underlying mechanisms behind our experimental
observations, we theoretically calculate the Andreev resonances and
their dependence on superconducting phase variables (see
section III in the
SI for details of our theoretical calculations).  We consider the proximity effect at a
disordered normal-metal–insulator–superconductor (NIS) interface.  At
the NIS interface, a spin-up electron from the
  disordered metal ($N$) can be Andreev reflected as a spin-down hole
while a Cooper pair is transmitted to $S$.  However, due to the low
contact transparency, a spin-up electron is highly likely to undergo
normal reflection back into $N$ and subsequently
return to the interface because of the localizing
  effect of disorder.  This process, involving multiple scattering events in the presence of disorder, can iterate until a Cooper pair is transmitted to $S$, and a hole is back-scattered to the normal region 
~\cite{Kulik1970}.  To describe the analogous
proximity-induced quartet mechanism in
  graphene-based MJJs, we employ a simple Random Phase Approximation
(RPA) in which an elementary kernel is iterated (see
Eq. 9 in the SI).  Considering the two-dimensional nature ($D=2$)
of graphene, we find that constructive interference of the proximity
effect emerges at fixed values of the superconducting
  phase difference for the two-terminal Josephson configuration, and
  at fixed values of the quartet phase in the three-terminal case. 

%%%%%%%%%%%%%%%%%%%%%%%%

Within this framework, a Cooper quartet resonance occurs when the
total phase accumulated through the transmission of two pairs into
$S_L$ via $S_R$, $S_B$ and $S_R$ via $S_L$, $S_B$,
becomes constructive (see
Fig. S6). The quartet phases correspond to the
sum of four Andreev reflection phase shifts at the SN interfaces,
i.e., $\chi_q=2\varphi_L-\varphi_R -\varphi_B$,
$-\varphi_L+2\varphi_R-\varphi_B$, and $-\varphi_L-\varphi_R
+2\varphi_B$. We calculate the nonlocal Andreev reflection using
Green's function formalism, where each reflection contributes a phase
shift of $\pi/2$.  From the constructive-interference condition in 2D,
we obtain the spectrum of Cooper-quartet resonances and the associated
energies (see Eq. 34 in the SI).

%In two dimensions, constructive interference of nonlocal Andreev
%processes fixes the quartet phase rather than the energy, implying
%the resonance condition holds along the same linear phase
%combinations (see section ~\ref{sec:2D} in SI). As a result, the
%quartet features appear as stationary lines in the
%$(\varphi_L,\varphi_R)$ plane and should persist across all energies
%when tuning the tunneling bias $V_t$. Details of the calculations are
%included in the SI.

The theoretically calculated energy-phase relation of quartet
resonances is consistent with our measurements, where we observe
resonance lines along $\chi_q$ in the $G$ map as shown in
Fig.~\ref{Ch6_DOS.jpg}. However, the bias voltage $V_t$ acts as an
energy filter that controls transmission through different conduction
channels. Energies near the gap edge correspond to channels with low
transmission, while energies toward the middle of the gap correspond
to more transparent, highly transmitting channels. Low transmission
favors the two-terminal DC Josephson effect, whereas higher-order
Andreev processes emerge at higher transmission. 
  This explains why the DC Josephson effect appears near the gap edge,
  while quartet resonances occur near mid-gap. These findings are in
  agreement with our experimental observation of quartet resonance
  around $V_t \approx \Delta/1.8 $ ($\approx$100 $\mu$V).

We investigate the ABS resonances in the regime where
  \( V_t \geq 2\Delta \) (see Figs.~S2). In
  this regime, the ABS energies coincide with the nonequilibrium Fermi
  surface at the energy \( \mu_N \) induced in
  the graphene region. The energy \( \mu_N \), approximated by \(
  \mu_N^* \sim \sqrt{(eV_t)^2 - (2\Delta)^2} \), can be substantially lower than the applied voltage \( V_t \) to the superconducting probe \( S_T \). Consequently, even in high bias regime \( V_t \geq 2\Delta \),
  the effective Fermi energy \( \mu_N \leq \Delta
  \) remains smaller than the gap, allowing spectroscopy of in-gap states (see
  Fig.~S7). As a result, we observe the
  characteristic signature of the quartet resonant family that
  reappears at \( V_t \approx 2.78\Delta \       \ (\approx500~\mu\text{V}) \) (see
  Fig.~S2(d)), consistent with the features
  previously identified in Fig.~\ref{Ch6_DOS.jpg}(b). Taken together,
  these observations demonstrate the robustness of quartet resonances
  against large bias voltages and highlight the versatility of our
  device for engineering high‑order correlated processes.

Figure~\ref{Ch6_DOS.jpg}(d) shows the
theoretically calculated tunneling conductance
map in the $(\varphi_L,\varphi_R)$ plane. The
  phenomenological model incorporates the two possible quartet
  configurations and calculates the conductance. The conductance map
  \(g_{0}(\varphi_{L},\varphi_{R},V_{t,\mathrm{eff}})\) is then
  computed on a discrete grid of phase values to produce the
  intersecting quartet resonance lines which qualitatively reproduces
  the key features of the quartet resonances observed
  experimentally. The details of the calculation is provided in section VI of the SI.

\begin{figure}[ht]
\centering
\includegraphics[width=10 cm]{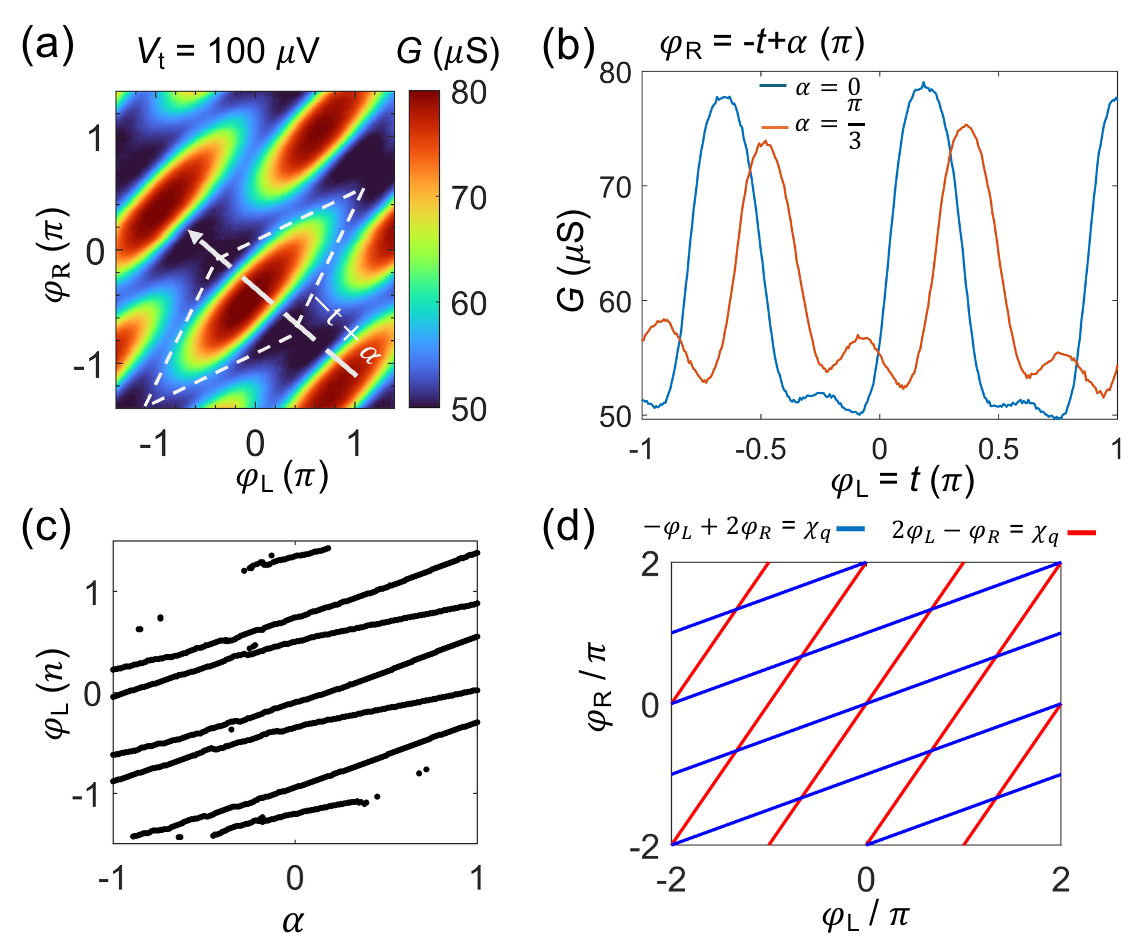}
 \caption{\label{Ch6_spegettie.jpg} \textbf{Hybridization of quartet
     modes}. (a) Tunneling conductance $G$ as a function of
   independent phases $\varphi_L$ and $\varphi_R$ at the bias voltage
   $ V_t$ = 100 $\mu$V. The colormap is obtained by varying $I_L$ and
   $I_R$ with a step sized of 0.25 $\mu$A and 1 $\mu$A,
   respectively. White dashed lines (diamond shape)
  indicate two modes of the quartet resonance with
     quartet phase variables $\chi_q = 2\varphi_L-\varphi_R$ and
     $\chi_q = -\varphi_L+2\varphi_R$. (b) Diagonal line cuts along
     $\varphi_L$ = $t$ and $\varphi_R$ = $-t+\alpha$ with the variable
     $\varphi_L$ = $t$ plotted on the x-axis and the tunneling
     conductance $G$ on the y-axis for $\alpha = 0$ (blue) and $\alpha
     = \pi/3$ (orange). Local minina of $G$ can be obtained at each
     value of $\alpha$.  (c) The scatter plot shows $\varphi_L$ ($n$)
     as a function of $\alpha$. The minima of $G$ and the
     corresponding phases $\varphi_L$ ($n$) is obtained by varying
     $\alpha~(\pi)$ (between -1 to 1) with a step size of $1/100$.  (d)
     Illustration of two families of Cooper quartet resonances with
     quartet phase variables $\chi_q = 2\varphi_L-\varphi_R$ and
     $\chi_q = -\varphi_L+2\varphi_R$ indicating  slope 2 (red) and slope 1/2 (blue),
     respectively.}
\end{figure}

Figure.~\ref{Ch6_spegettie.jpg}(a) shows the differential conductance
$G$ as a function of $\varphi_L$ and $\varphi_R$ at the bias voltage $
V_t$ = 100 $\mu$V. The white dashed lines indicate the two
quartet-resonance branches forming a diamond like pattern. To
experimentally track the behavior of these two branches, we extract
diagonal line cuts from Fig.~\ref{Ch6_spegettie.jpg}(a) along
$\varphi_L$ = $t$ and $\varphi_R$ =
$-t+\alpha$. {Figure~\ref{Ch6_spegettie.jpg}(b) plots
  line cuts of $G$ as a function of the variable $t$ for $\alpha = 0$
  (blue) and $\alpha = \pi/3$ (orange) . We can obtain line cuts
  similar to Fig.~\ref{Ch6_spegettie.jpg}(b) by varying $\alpha~(\pi)$
  between -1 to 1 with a step size of $1/100$. For each trace (using a
  Gaussian smoothing), we identify the local minima in $G$ and record
  the corresponding phases $\varphi_L$ ($n$) at the values of
  $\alpha$. These points are plotted in Fig.~\ref{Ch6_spegettie.jpg}(c), 
  yielding a series of continuous curves that track quartet
branches as $\alpha$ is varied. Notably, these curves exhibit robust
continuity and show avoided crossing between the branches, indicating
hybridization of quartet modes.

Figure~\ref{Ch6_spegettie.jpg}(d) shows theoretically calculated two
quartet-resonance branches with quartet phase variables $\chi_q =
2\varphi_L-\varphi_R$ and $\chi_q = -\varphi_L+2\varphi_R$ indicating
slope 2 (red) and slope 1/2 (blue), respectively. We see that the
resonant interaction of Cooper quartets in
Fig.~\ref{Ch6_spegettie.jpg}(c) contrasts with the
assumption of independent quartet spectra that
  underlies the theoretical calculations. In contrast to experiment,
the calculated quartet spectra show direct crossings between two
quartet branches. At these crossings, bistability emerges between
distinct quartet states. In one- and three-dimensional systems, such
intersections would normally lead to level repulsion, causing the DC
quartet current (proportional to $\delta E/\delta\varphi$) to vanish
at the avoided crossing. However, in large-scale
two-dimensional systems the quartet spectrum forms a continuum at
fixed superconducting phases. But we argue that,
if the two quartet branches crossed without
hybridization, the coexistence of degenerate states would generate
partition noise in the DC quartet supercurrent, which contradicts the
well-established result that Josephson currents between BCS
superconductors are noiseless at zero temperature
\cite{Melinnoise2016}.  This apparent
contradiction is resolved by recognizing that the quartet current must
vanish at the crossing points of the two branches. The
theoretical Fig.~\ref{Ch6_spegettie.jpg}(d) is deduced from a multiterminal
generalization of long highly transparent two-terminal Josephson junctions \cite{Kulik1970}, where the ABS energies are free to cross at the phase differences of $0$ and $\pi$. Including the realistic backscattering produces avoided crossings in the two-terminal ABS spectra, and hybridization between the ABS branches \cite{Bagwell1992}. Thus, theory predicts a vanishing quartet
current at the crossing points, consistent with the avoided crossings
(hybridization) observed in the experimental $G$ maps.

Although our analysis focuses on the microscopic origin of Cooper
quartets, the structure of the resonances naturally invites a broader
topological interpretation. Because the superconducting phase is
defined modulo \(2\pi\), the control parameters
\((\varphi_{L}, \varphi_{R})\) form a compact
two-dimensional torus that acts as a synthetic Brillouin zone for the
Andreev spectrum. The two observed distinct
  branches of quartet resonances follow the relations
\(2\varphi_{L} - \varphi_{R} = \mathrm{const.}\) and
  \(-\varphi_{L} + 2\varphi_{R} = \mathrm{const}\). These relations
correspond to straight, quantized winding trajectories on this
toroidal phase space. These winding conditions of the quartets echo
classical topological classifications of closed trajectories,
analogous to early knot theory \cite{Haken1962topologyknot}. In
multiterminal Josephson junctions, integer combinations of
superconducting phases play a similar role and may be used to organize
higher-order Cooper multiplets such as quartets, sextets, and beyond.

Our experiment demonstrates a key departure from this classical picture. At the intersection of the two winding families, the resonances do not cross freely, as would be expected from topologically distinct classical trajectories. Instead, we observe clear avoided crossings. This behavior reflects coherent hybridization between Andreev bound states belonging to different winding sectors and shows that the Andreev spectrum cannot be understood by classical winding arguments alone. The avoided crossings therefore serve as direct signatures of the quantum nature of the Andreev bands on the torus of the
    superconducting phase variables. The winding conditions determine where the resonances appear, while quantum coherence controls how these resonances interact. Classical winding topology alone cannot account for this hybridization, its resolution requires the full quantum mechanical structure of the Andreev spectrum. 

Taken together, these results suggest that multiterminal Josephson junctions naturally host two distinct forms of topology. One is classical, defined by integer phase-winding sectors that classify multiplet processes. The other is quantum, encoded in the Andreev band structure on the phase torus, which can in principle carry Berry curvature \cite{Peraltaberrycurvature}, Chern number \cite{Nazarov2019}, or isolated singularities in higher dimensional phase spaces. The observed hybridization of quartet modes demonstrates that these two distinct forms of topology can interact. Extending this spectroscopic approach to four-terminal junctions, where the phase space becomes three dimensional, offers a promising route to explore whether these two topological structures remain separate or combine to produce new forms of topology unique to superconducting circuits with phase control.

\section*{Conclusion}
In summary, we have performed a tomographic study of Andreev bound states in a graphene-based three-terminal Josephson junction using phase-resolved tunneling spectroscopy.  Independent control of the left and right phases through a double-SQUID geometry enabled direct mapping of the synthetic Andreev band structure. By tuning the bias voltage $V_t$, we accessed higher-order Andreev processes and identified resonances that satisfy the quartet phase relations.  Because the phase space forms a compact torus, these resonances trace quantized winding trajectories associated with distinct quartet modes. At the intersections of these trajectories, we observe avoided crossings,  signaling coherent hybridization between modes belonging to different winding sectors. A microscopic model  reproduces the phase structure and explains the avoided crossings.

When $V_t > 2\Delta/e$, we showed that quartet patterns re-emerge, owing to relaxation of injected quasiparticles toward an effective nonequilibrium chemical potential. By establishing a direct, phase-resolved link between tunable multiterminal geometries and high-order Cooper-pair transport, our results provide a quantitative benchmark for the detection of Cooper quartets and demonstrate a practical path toward engineering synthetic Andreev band structures with non-trivial topology. This spectroscopic approach can now be extended to four-terminal junctions, where higher-order Copper multiplets (sextets) \cite{ebert2025sextets}, Weyl ,  and additional topological phases are predicted \cite{xie2018weyl, burkov2011weyl}. Graphene multiterminal junctions therefore emerge as a powerful and versatile platform for manipulating correlated superconducting states, and for  realizing quantum phase design in higher-dimensional Andreev crystals.

\section*{Materials and Methods}
\label{method}

\subsection*{Device Fabrication}Graphene and hexagonal boron nitride (hBN) flakes were mechanically exfoliated using adhesive tape. The
heterostructures were assembled using a standard dry-transfer technique, in
which graphene was encapsulated between hBN layers with a hBN thickness of
approximately 35~nm each. The device channel was defined by electron-beam
lithography followed by reactive ion etching using a CHF$_3$/O$_2$ plasma.
Electrical contacts were fabricated by electron-beam evaporation onto freshly
etched graphene edges. Ti (10~nm)/Al (80~nm) were deposited to form edge
contacts and flux lines. A superconducting tunnel probe, consisting
of an 80~nm Al layer without an adhesion layer, was deposited using BN--TiB$_2$
crucibles at a rate of 0.4~\AA~s$^{-1}$. During deposition, the chamber pressure
was maintained at $1 \times 10^{-7}$~mtorr.
\subsection*{Electrical Transport}
All the electrical transport measurements were performed inside a Bluefors LD250 dilution fridge, equipped with QDevil filtered DC lines. A dc bias voltage $V_t$ was applied to the tunnel probe with a small signal ac voltage $V_{\text{ac}}\approx 5~\mu$V, and the tunneling voltage ($dV_t$) was measured between $S_T$ and node $D$ using a lock-in amplifier. Another lock-in amplifier was used to measure the tunneling current ($dI_t$). The tunneling current $dI_t$ was amplified via a current to voltage preamplifier whose voltage was subsequently measured using another lock-in amplifier. The resulted differential tunneling conductance is obtained as $G = {dI_t}/{dV_t}$.
\section*{Acknowledgements}
\subsection*{Funding} M.K. and A.S.R. acknowledge funding from the the US National Science Foundation under award DMR 2415756. Device fabrication was supported through the Pennsylvania State University Materials Research Science and Engineering Center supported by the US National Science Foundation (DMR 2011839). R.M. acknowledges the financial support from the SUPRADEVMAT
International Project between the French CNRS-Grenoble and the German
KIT-Karlsruhe.
\subsection*{Data Availability}
All data supporting the conclusions in the paper are present in the paper and/or the Supplementary Materials.
%%%%%%%%%%%%%%%% REFERENCES %%%%%%%%%%%%%%%

\clearpage % Clear all remaining figures and tables then start a new page

% The list of references goes after the main text and before the acknowledgements
% When preparing an initial submission, we recommend you use BibTeX, like this:
%
\bibliography{References} % for a file named science_template.bib
\bibliographystyle{sciencemag}

%%%%%%%%%%%%%%%% START OF SUPPLEMENT %%%%%%%%%%%%%%%
\clearpage
\newpage

\renewcommand\thefigure{S\arabic{figure}}    
\setcounter{figure}{0}

\centerline{\textbf{\Large{Supporting Information}}}
\AtBeginDocument{\renewcommand{\figurename}{\textbf{Figure}}}
\renewcommand\thefigure{\textbf{S}\arabic{figure}}    
\setcounter{figure}{0}  
\renewcommand{\thesection}{\Roman{section}}
\section{Tunneling Spectroscopy}\label{gap}
We have fabricated two devices that display distinct tunneling characteristics arising from variations in the tunneling interface during nanofabrication. Device 1 exhibits signatures of a soft superconducting gap, whereas device 2 shows signatures of a hard superconducting gap at $T$ = 50 mk, as illustrated in Fig.~\ref{Ch6_gap.jpg}.
\begin{figure}[hb]
\centering
\includegraphics[width=14 cm]{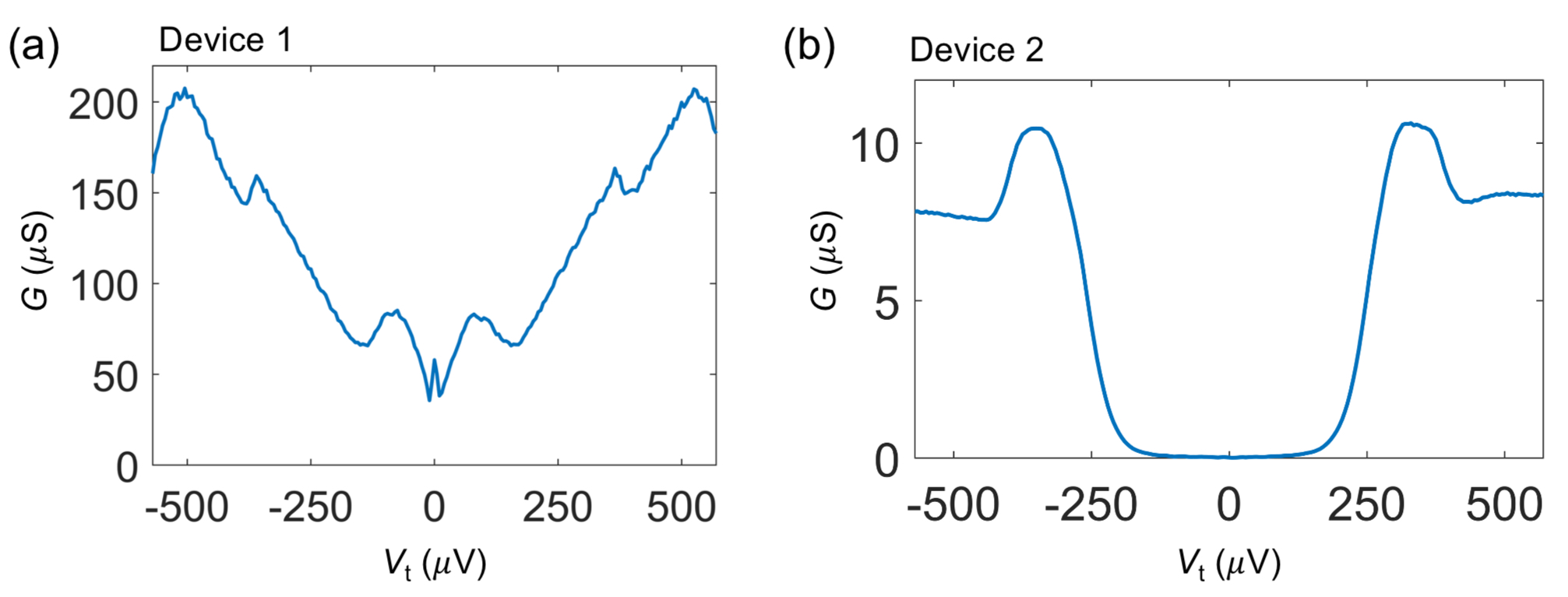}
 \caption{\label{Ch6_gap.jpg} \textbf{Tunneling gap.} (a) Tunneling conductance $G$ as a function of the bias voltage $V_t$ in device 1. (b) Tunneling conductance $G$ as a function of the bias voltage $V_t$ for device 2. }
\end{figure}

\section{Tunneling conductance as a function of $\varphi_L$, $\varphi_R$ for $ V_t > \Delta$}

Figures~\ref{Ch6_conductance.jpg} and ~\ref{Ch6_device2conductance.jpg} shows the tunneling conductance $G$ as a function of independent phases $\varphi_L$ and $\varphi_R$ at fixed bias voltage $ V_t$’s ($V_t > \Delta$) for devices 1 and 2, respectively. In device 1, the characteristic signature of the quartet-resonant family reappears at \( V_t = 500~\mu\text{V} \), consistent with our observations at \( V_t = 100~\mu\text{V} \) in the main text. Similarly, in device 2,  two families of quartet resonances emerge as the characteristic diamond-shaped pattern at \( V_t = 350~\mu\text{V} \) as shown in Fig.~\ref{Ch6_device2conductance.jpg} (c).

\begin{figure}[ht]
\centering
\includegraphics[width=10 cm]{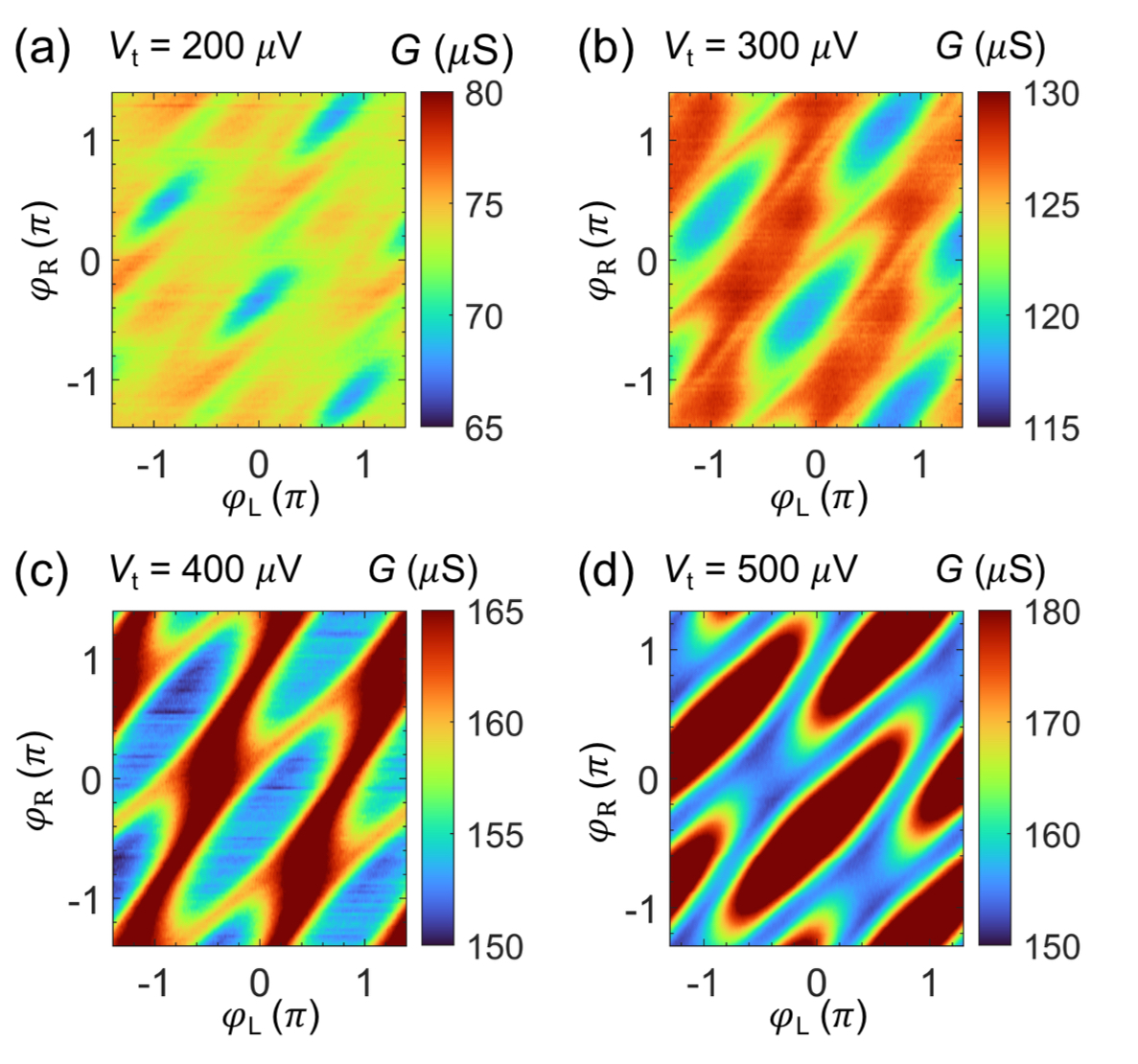}
 \caption{\label{Ch6_conductance.jpg} \textbf{Tunneling spectroscopy in three-terminal JJ at $V_t \geq \Delta$.} (a-d) Tunneling conductance $G$ as a function of independent phases $\varphi_L$ and $\varphi_R$ governed by fluxes $\Phi_L$ and $\Phi_R$ at constant $ V_t$’s ($V_t \geq \Delta$) for device 1. The colormaps are obtained by varying $I_L$ and $I_R$ with a step sized of 0.25 $\mu$A and 1 $\mu$A respectively. The colormaps are measured at $T$ = 200 mK and back gate voltage $V_g$ = 0 V. }
\end{figure}
\newpage

\begin{figure}[t]
\centering
\includegraphics[width=14 cm]{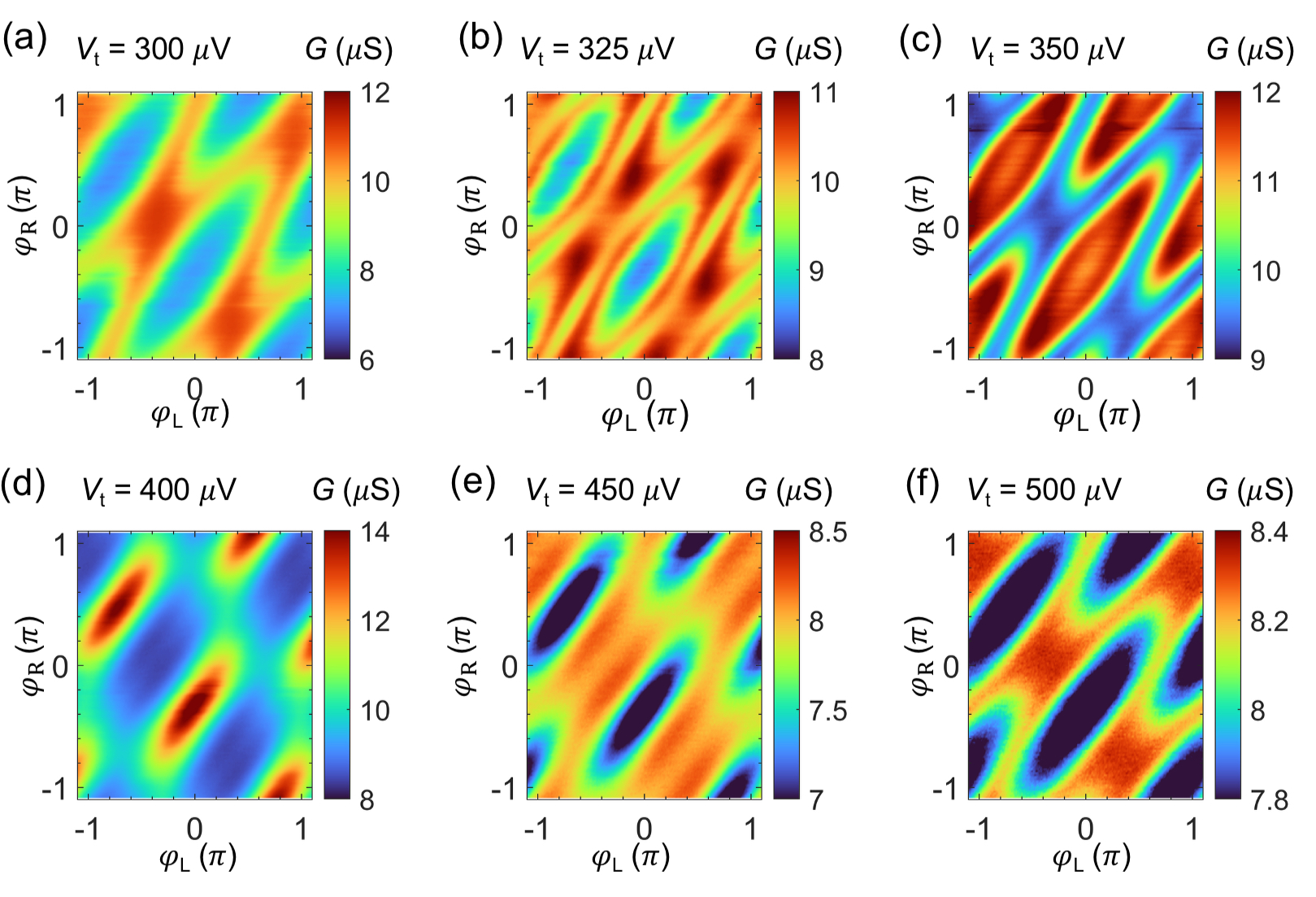}
 \caption{\label{Ch6_device2conductance.jpg} \textbf{Tunneling spectroscopy for device 2 at $V_t \geq \Delta$.} (a-f) Tunneling conductance $G$ as a function of phases $\varphi_L$ and $\varphi_R$ at constant $ V_t$’s ($V_t > \Delta$) for device 2. The colormaps are obtained by varying $I_L$ and $I_R$ with a step sized of 0.5 $\mu$A and 1 $\mu$A, respectively. The colormaps are measured at $T$ = 50 mK and back gate voltage $V_g$ = 0 V.}
\end{figure}

\section{Multipair Andreev resonances in 1D and 3D}\label{sec:1D-3D}
In this section, we calculate the spectrum of resonances:
the Cooper pairs and the Cooper quartets.
\subsubsection{Cooper pair resonances}
In this subsection, we calculate the spectrum of the Cooper-pair resonances. To describe proximity-induced Cooper-pair and multipair correlations in the graphene region, we employ the Dyson equation formalism, which provides a systematic framework to account for tunneling processes between the normal region and the superconducting leads to all orders. The general form of the Dyson equations is the following: 
\label{sec:Cooper-pairs}

\begin{equation}
  \label{eq:Dyson0}
  \hat{G}_{D,D}=\hat{g}_{D,D}
  + \hat{g}_{D,D}\hat{\Sigma}_{D,S} \hat{G}_{S,D}
  ,
\end{equation}
where $\hat{G}$ and $\hat{g}$ are the fully dressed and bare Nambu
Green's function and $\hat{\Sigma}$ is the hopping self-energy. The
subscript ``D'' refers to the ``dot'', i.e. the piece of graphene, and
the subscript ``S'' refers to the superconducting leads. The symbol
``D'' is used for a general rectangular-shaped 2D metal. Using:
\begin{equation}
  \label{eq:Dyson0-bis}
  \hat{G}_{S,D}=
 \hat{g}_{S,S}\hat{\Sigma}_{S,D} \hat{G}_{D,D}
\end{equation}
and combining Eqs.~(\ref{eq:Dyson0}) with~(\ref{eq:Dyson0-bis}) lead
to
\begin{equation}
  \label{eq:Dyson1}
  \hat{G}_{D,D}=\hat{g}_{D,D}
  + \hat{g}_{D,D}\hat{\Sigma}_{D,S} \hat{g}_{S,S} \hat{\Sigma}_{S,D} \hat{G}_{D,D}
  ,
\end{equation}
This equation represents a closed system of linear equations for the fully dressed
Nambu Green's function $\hat{G}_{D,D}$. At the next iteration,
Eq.~(\ref{eq:Dyson1}) can be rewritten as:
\begin{equation}
  \label{eq:Dyson1-bis}
  \hat{G}_{D,D}=\hat{g}_{D,D} + \hat{g}_{D,D}\hat{\Sigma}_{D,S}
  \hat{g}_{S,S} \hat{\Sigma}_{S,D} \hat{g}_{D,D}+
  \hat{g}_{D,D}\hat{\Sigma}_{D,S} \hat{g}_{S,S} \hat{\Sigma}_{S,D}
  \hat{g}_{D,D}\hat{\Sigma}_{D,S} \hat{g}_{S,S} \hat{\Sigma}_{S,D}
  \hat{G}_{D,D}
  .
\end{equation}
Eqs.~(\ref{eq:Dyson1}) and (\ref{eq:Dyson1-bis}) are exact and 
are equivalent to each other.

We now apply an approximation in which the Green's functions are
grouped pairwise, and the resulting set of paired Green functions is
summed to infinite order in the form of a matrix geometric
series. This random phase approximation (RPA) involves averaging, in
real space, the modes (i.e. the pairs of Green's functions) over
oscillations occurring on the scale of the Fermi wave-length
$\lambda_F$, in a manner reminiscent of the formation of
two-particle interference modes
known as diffusons and
  Cooperons in disordered metals. In a ballistic 2D metal, the
Cooper-pair modes are also known as {\it Andreev tubes}; see
Refs.~\cite{Meier2016,Kraft2018V2}. We denote the corresponding
averaging by the double bracket $\langle \langle ... \rangle \rangle$:
\begin{equation}
  \label{eq:Dyson2}
  \langle \langle\hat{G}_{D,D}\rangle \rangle^{(pairs)}=
  \langle\langle
  \hat{g}_{D,D}\rangle \rangle + \langle
  \langle\hat{g}_{D,D}\hat{\Sigma}_{D,S} \hat{g}_{S,S}
  \hat{\Sigma}_{S,D} \hat{g}_{D,D}\rangle \rangle+\langle \langle
  \hat{g}_{D,D}\hat{\Sigma}_{D,S} \hat{g}_{S,S} \hat{\Sigma}_{S,D}
  \hat{g}_{D,D}\hat{\Sigma}_{D,S} \hat{g}_{S,S}
  \hat{\Sigma}_{S,D}\rangle \rangle \langle
  \langle\hat{G}_{D,D}\rangle \rangle^{(pairs)},
\end{equation}
where we implement a kind of decoupling in the last term. Within
this RPA, Eq.~(\ref{eq:Dyson2}) is rewritten as
\begin{equation}
  \label{eq:Dyson2-ter}
  \langle \langle\hat{G}_{D,D}\rangle \rangle^{(pairs)}= \left[\hat{I}-\langle
    \langle \hat{g}_{D,D} \hat{\Gamma}_{D,D}
    \hat{g}_{D,D}\hat{\Gamma}_{D,D} \rangle \rangle\right]^{-1} \times
  \left[ \langle \langle\hat{g}_{D,D}\rangle \rangle + \langle
    \langle\hat{g}_{D,D}\hat{\Gamma}_{D,D} \hat{g}_{D,D}\rangle
    \rangle \right],
\end{equation}
where $\hat{\Gamma}_{D,D} = \hat{\Sigma}_{D,S}
\hat{g}_{S,S} \hat{\Sigma}_{S,D}$, and  $\langle\langle
\hat{g}_{D,D} \rangle\rangle$ is local in space because it involves
averaging over a single bare Green's function.

Next, we introduce the notations $\hat{g}_{D,D}^{loc}$ and
$\hat{g}_{D,D}^{non\,loc}$ for the local and the nonlocal Green's
functions of the 2D metal, respectively. We also work within the
phenomenological framework of the large-gap approximation
\cite{Zazunov2003,Meng2009,Melin2021,Klees2020,Melin2024a}, which
allows us to base the physical discussion on the simplest
approximation. In this approximation, the normal bare Nambu-diagonal
superconducting Green's functions $g_{S,S}^{1,1}=g_{S,S}^{2,2}=0$ are
taken to be vanishingly small. Under these conditions,
 Eq.~(\ref{eq:Dyson2-ter}) takes the following
  form:
\begin{eqnarray}
  \label{eq:Dyson3}
  \langle \langle\hat{G}_{D,D}\rangle \rangle^{(pairs)} &=&
  \hat{D}_0^{4\times 4} \left[ \langle \langle\hat{g}_{D,D}\rangle
    \rangle + \langle \langle\hat{g}_{D,D}\hat{\Gamma}_{D,D}
    \hat{g}_{D,D}\rangle \rangle \right] ,
\end{eqnarray}
where $\hat{D}_0^{4\times 4}$ is expressed as a $4\times 4$ matrix
in the basis of the electron and hole Nambu labels, and the
tight-binding sites $\alpha$ and $\beta$ at both contacts:
\begin{eqnarray}
  \hat{D}_0^{4\times 4}=
  \left(\begin{array}{cccc}
    1-\hat{K}^{1,1}_{D_\alpha,D_\alpha}&-\hat{K}^{1,1}_{D_\alpha,D_\beta}&-\hat{K}_{D_\alpha,D_\alpha}^{1,2}&-\hat{K}_{D_\alpha,D_\beta}^{1,2}\\
    -\hat{K}^{1,1}_{D_\beta,D_\alpha}&1-\hat{K}^{1,1}_{D_\beta,D_\beta}&-\hat{K}_{D_\beta,D_\alpha}^{1,2}&-\hat{K}_{D_\beta,D_\beta}^{1,2}\\
    -\hat{K}^{2,1}_{D_\alpha,D_\alpha}&-\hat{K}^{2,1}_{D_\alpha,D_\beta}&1-\hat{K}_{D_\alpha,D_\alpha}^{2,2}&-\hat{K}_{D_\alpha,D_\beta}^{2,2}\\
    -\hat{K}^{2,1}_{D_\beta,D_\alpha}&-\hat{K}^{2,1}_{D_\beta,D_\beta}&-\hat{K}_{D_\beta,D_\alpha}^{2,2}&1-\hat{K}_{D_\beta,D_\beta}^{2,2}
  \end{array}\right)
  ,
\end{eqnarray}
where the kernel $\hat{K}$ is:
\begin{equation}
  \label{eq:kernel}
  \hat{K}=\hat{g}_{D,D}\hat{\Gamma}_{D,D}\hat{g}_{D,D}\hat{\Gamma}_{D,D}
  .
\end{equation}

%%%%%%%%%%%%%%%%%%%%%%%%%%%%%%%%%%%%%%%%%%%%%%%%%%%%%%%%%%%%%%%%%%%%%%%
\begin{figure}
\centering

    \includegraphics[width=9 cm]{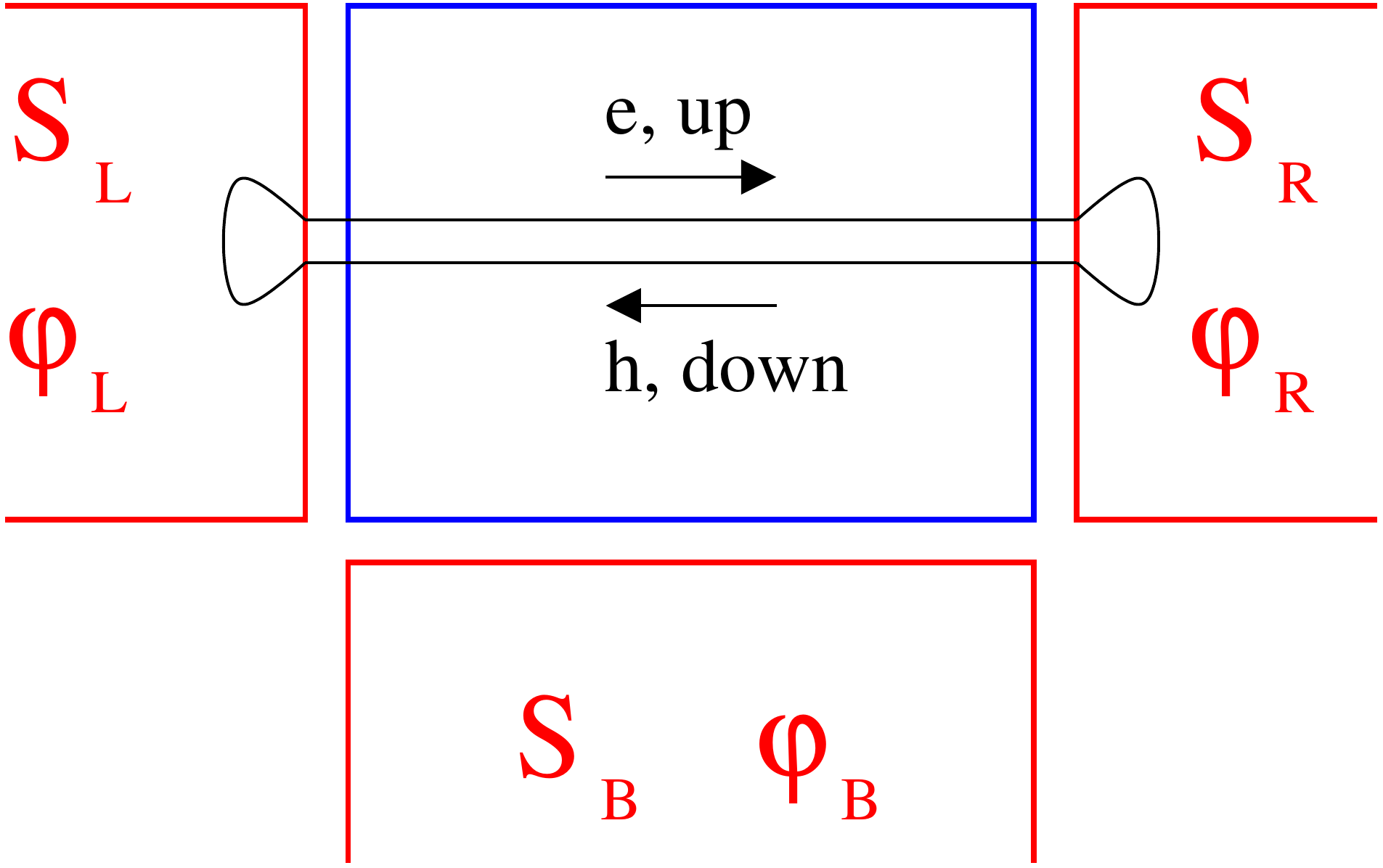}
    
\caption{A diagram representing how a 1D Andreev tube propagating
  through the 2D metal connects two Andreev and inverse-Andreev
  reflection processes at the right and left
  superconductors $S_R$ and $S_L$, respectively.
\label{fig:Cooper-pairs}
}

\end{figure}

%%%%%%%%%%%%%%%%%%%%%%%%%%%%%%%%%%%%%%%%%%%%%%%%%%%%%%%%%%%%%%%%%%%%%%%%%%%

Each of the $\hat{\Gamma}$ matrices necessarily produces either an
electron-hole Andreev reflection or a hole-electron inverse Andreev
reflection within the large-gap approximation. The matrix $\hat{K}$
contains a pair of $\hat{\Gamma}$ terms, and its off-diagonal Nambu
elements are vanishingly small in this approximation,
i.e. $\hat{K}^{1,2}=\hat{K}^{2,1}=0$. Moreover, averaging over
$\lambda_F$ causes products of local and nonlocal Green's functions to
vanish, leading to a negligible contribution of $\langle\langle
\hat{K}_{D_\alpha,D_\beta} \rangle\rangle$ for $\beta\ne\alpha$. A
nonlocal component of the kernel $\hat{K}$ is schematically
illustrated in Fig.~\ref{fig:Cooper-pairs}. We conclude that the
matrix $\langle\langle \hat{D}_0^{4\times 4} \rangle\rangle$ becomes
diagonal after averaging over rapid Fermi-phase
  oscillations on the scale of the Fermi
wave-length $\lambda_F$, see the corresponding
  $\exp(\pm i k_F R_{\alpha,\beta})$ terms in the forthcoming
  Eqs.~(\ref{eq:g-1D-1})-(\ref{eq:g-3D-2}):
\begin{eqnarray}
  \label{eq:matrix}
  \langle\langle\hat{D}_0^{4\times 4}\rangle\rangle=
  \left(\begin{array}{cccc}
    1-\langle\langle\hat{K}^{1,1}_{D_\alpha,D_\alpha}\rangle\rangle & 0 & 0 & 0\\
    0 & 1-\langle\langle\hat{K}^{1,1}_{D_\beta,D_\beta}\rangle\rangle & 0 & 0\\
    0 & 0 & 1-\langle\langle\hat{K}_{D_\alpha,D_\alpha}^{2,2}\rangle\rangle& 0\\
    0 & 0 & 0 &1-\langle\langle\hat{K}_{D_\beta,D_\beta}^{2,2}\rangle\rangle
  \end{array}\right)
  .
\end{eqnarray}
Taking into account the first entry of Eq.~(\ref{eq:matrix}), we find
\begin{equation}
  1-\langle\langle \hat{K}_{D_\alpha,D_\alpha}^{1,1} \rangle\rangle \simeq
  1- \hat{K}_{D_\alpha,D_\alpha}^{1,1,loc} -  \langle\langle \hat{K}_{D_\alpha,D_\alpha}^{1,1,non\,loc} \rangle\rangle
  ,
\end{equation}
with
\begin{eqnarray}
  \hat{K}_{D_\alpha,D_\alpha}^{1,1,loc}&\simeq& g_{\alpha,\alpha}^{1,1}
  \Gamma_{\alpha,\alpha}^{1,2} g_{\alpha,\alpha}^{2,2}
  \Gamma_{\alpha,\alpha}^{2,1}\\ \langle\langle
  \hat{K}_{D_\alpha,D_\alpha}^{1,1,non\,loc}\rangle\rangle &\simeq&
  \langle\langle
  g_{\alpha,\beta}^{1,1} \Gamma_{\beta,\beta}^{1,2}
  g_{\beta,\alpha}^{2,2} \Gamma_{\alpha,\alpha}^{2,1}
  \rangle\rangle
  \label{eq:K-bis}
  ,
\end{eqnarray}
where $g_{\alpha,\alpha}^{1,1}=g_{\alpha,\alpha}^{2,2} \simeq i / W$,
$\Gamma_{\alpha,\alpha}^{1,2} \simeq -(\Sigma_{D_\alpha,S_\alpha}^2/W)
\exp(i\varphi_\alpha)$, $\Gamma_{\alpha,\alpha}^{2,1} \simeq
-(\Sigma_{D_\alpha,S_\alpha}^2/W) \exp(-i\varphi_\alpha)$, and $W$ is
the largest energy scale set by the bandwidth of the
  normal region dispersion relation. We conclude that $\hat{K}_{D_\alpha,D_\alpha}^{1,1,loc}
\simeq - \Sigma_{D_\alpha,S_\alpha}^4 / W^4$ has a vanishingly small
imaginary part, a result that will be used in the following
calculations.

The inverse of each diagonal matrix element of Eq.~(\ref{eq:matrix})
generically takes the following form:
\begin{equation}
  \label{eq:series}
  \frac{1}{1-K}=\frac{1}{1-K_{loc}-K_{non\,loc}}=
  \frac{1}{1-K_{loc}}\sum_{n=0}^{+\infty} p^n
  ,
\end{equation}
whre $n$ is an integer,  $p=K_{non\,loc}/(1-K_{loc})$, and
``loc'' and ``non loc'' in the subscript refer to the local and
nonlocal components of the matrix $\hat{K}$. Constructive interference
is possible only in the presence of a vanishingly small dephasing in
the variable $p$. This type of interference is reminiscent of the
proximity effect at a normal metal-insulator-superconductor ($NIS$)
interface in the presence of nonmagnetic impurities in the normal region
$N$ \cite{HN1,HN2,Beenakker}. In the proximity effect, disorder-induced localization
leads to multiple backscattering events at the $NIS$
interface and, at low energies, enables constructive interference
between the corresponding semiclassical trajectories of Andreev
pairs.

Now, we provide the expression of the nonlocal Green's functions
$\hat{g}_{non\,loc}$ which will subsequently be inserted into
Eq.~(\ref{eq:K-bis}) and Eq.~(\ref{eq:series}). Following a standard
procedure, these Green's functions are obtained by integrating over
the wave-vector $k$, which is treated as a continuous variable. In one
dimension (1D), this yields the following expression for the nonlocal
Green's function:
\begin{eqnarray}
  \label{eq:g-1D-1}
  g_{\alpha,\beta,1,1}^{1D} \simeq \frac{1}{W}
  \exp{\left[-i\left(k_F+\frac{\omega}{v_F}\right)
      R_{\alpha,\beta}\right]}\\ g_{\alpha,\beta,2,2}^{1D} \simeq -\frac{1}{W}
  \exp{\left[i\left(k_F-\frac{\omega}{v_F}\right) R_{\alpha,\beta}\right]} ,
  \label{eq:g-1D-2}
\end{eqnarray}
where $R_{\alpha,\beta}$ is the separation between the tight-binding
sites $\alpha$ and $\beta$. In three dimensions (3D), we obtain a
similar expression as Eqs.~(\ref{eq:g-1D-1})-(\ref{eq:g-1D-2}) in 1D,
but with a geometrical prefactor:
\begin{eqnarray}
  \label{eq:g-3D-1}
  g_{\alpha,\beta,1,1}^{3D} \simeq \frac{1}{W k_F R_{\alpha,\beta}}
  \exp{\left[-i\left(k_F+\frac{\omega}{v_F}\right)
      R_{\alpha,\beta}\right]}\\ g_{\alpha,\beta,2,2}^{3D} \simeq -\frac{1}{W k_F R_{\alpha,\beta}}
  \exp{\left[i\left(k_F-\frac{\omega}{v_F}\right) R_{\alpha,\beta}\right]}
  \label{eq:g-3D-2}
  .
\end{eqnarray}
In Eqs.~(\ref{eq:g-3D-1})-(\ref{eq:g-3D-2}), we approximated the
slowly-varying $(k_F\pm \omega/v_F)R_{\alpha,\beta}$ in the
denominator of the geometrical prefactor by $k_F
R_{\alpha,\beta}$.

By combining these orbital phase variables with the superconducting phases
associated with  electron-hole Andreev or hole-electron
inverse-Andreev conversions in the superconductors, we
find the following condition for full constructive interference:
\begin{equation}
  \label{eq:condition1}
\frac{2\omega R_{\alpha,\beta}}{\hbar v_F}
\pm\varphi_L\mp\varphi_R+\pi=2\pi n ,
\end{equation}
where $n$ is an integer. The $\pi$-shift in
Eq.~(\ref{eq:condition1}) originates from the minus signs in
Eqs.~(\ref{eq:g-1D-2}) and ~(\ref{eq:g-3D-2}), which are arising from
the pair of Andreev reflections contained in
$\hat{K}_{non\,loc}$. Each of these reflections contributes a phase
shift of $\pi/2$ when the energy is small compared to the
superconducting gap\cite{blonder1982transition}. Therefore, resonances whose real parts
coincide with the energies of the Kulik spectrum are generated
\cite{Kulik1970,Ishii1970,Bagwell1992}:
\begin{equation}
  \label{eq:E-pairs}
  E_n^{(pairs)}=\frac{\hbar v_F}{2 R_{\alpha,\beta}}\left[2\pi
    \left(n+\frac{1}{2}\right) \mp \chi_{p}\right] ,
\end{equation}
where $\chi_{p}=\varphi_L-\varphi_R$ is the
phase difference between the superconducting leads.
%%%%%%%%%%%%%%%%%%%%%%%%%%%%%%%%%%%%%%%%%%%%%%%%%%%%%%%%%%%%%%%%%%%%%%%
\begin{figure}
    \centering

    \includegraphics[width=9 cm]{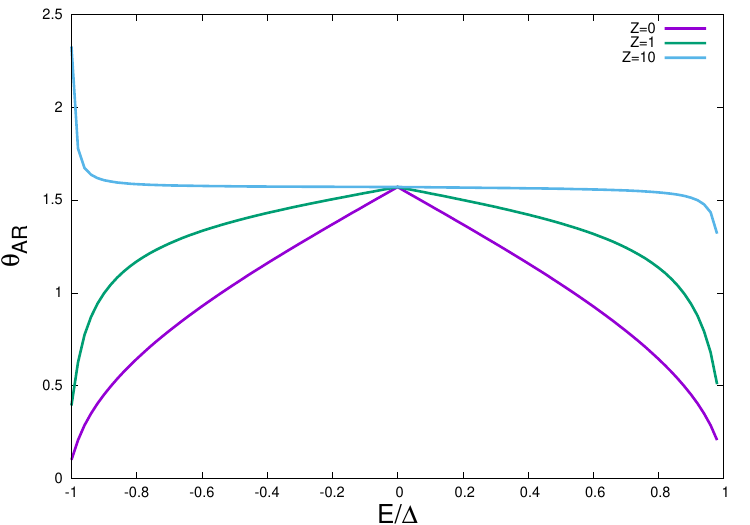}
    
    \caption{The phase shift $\theta_{AR}$ of a single Andreev
      reflection, calculated from Ref.~\cite{blonder1982transition}, and plotted
      as a function of the energy $E$ for the values $Z=0,\,1,\,10$.
\label{fig:theta}
}
\end{figure}
%%%%%%%%%%%%%%%%%%%%%%%%%%%%%%%%%%%%%%%%%%%%%%%%%%%%%%%%%%%%%%%%%%%
Fig.~\ref{fig:theta} shows the phase shift $\theta_{AR}$ of a single
Andreev reflection process at an arbitrary energy $E$ within the
superconducting gap $\Delta$, for Blonder–Tinkham–Klapwijk (BTK)
parameter values $Z=0,\,1,\,10$ \cite{blonder1982transition}. The
parameter $Z$ characterizes the strength of the repulsive interfacial
scattering potential, with high and low interface transparencies
corresponding to $Z\ll 1$ and $Z\gg 1$, respectively.  The phase shift
$\theta_{AR}$ is obtained from $\theta_{AR}=\arg(a)$, where the
Andreev reflection amplitude is given by
Eq.~(A11a) in Ref.~\cite{blonder1982transition}
\begin{equation}
  \label{eq:a}
  a=\frac{u_0v_0}{\gamma}
  ,
\end{equation}
The coherence factors are
denoted by
\begin{equation}
  u_0^2=\frac{1}{2}\left[1+\frac{\sqrt{E^2-\Delta^2}}{E}\right]
  .
\end{equation}
and $v_0^2=1-u_0^2$. The BTK notation $\gamma=u_0^2 +(u_0^2-v_0^2)Z^2$
was used in Eq.~(\ref{eq:a}).

By combining the phase shift $\theta_{AR}$ associated with a single
NIS interface with Eq.~(\ref{eq:E-pairs}) for the spectrum of a
two-terminal long SNS junction, we obtain the following
phenomenological expression for the ABS spectrum:
\begin{equation}
  \label{eq:theta-pairs}
  E_n^{(pairs)}=\frac{\hbar v_F}{2 R_{\alpha,\beta}}\left[2\pi n
    +2\theta_{AR} \mp \chi_{p}\right]
  .
\end{equation}

\subsubsection{Cooper quartet resonances}
\label{sec:Cooper-quartets}

We now calculate the spectrum of the Cooper quartet
resonances by extending the approximations to higher-order processes. 
Specifically, we expand
the Dyson Eq.~(\ref{eq:Dyson1-bis}) as follows:
\begin{eqnarray}
  \label{eq:Dyson-A}
  \hat{G}_{D,D}&=&\hat{g}_{D,D}\\ &+& \hat{g}_{D,D} \hat{\Gamma}_{D,D}
  \hat{g}_{D,D}\\&+& \hat{g}_{D,D}\hat{\Gamma}_{D,D}
  \hat{g}_{D,D}\hat{\Gamma}_{D,D}\hat{g}_{D,D}\\&+&
  \hat{g}_{D,D}\hat{\Gamma}_{D,D} \hat{g}_{D,D}\hat{\Gamma}_{D,D}
  \hat{g}_{D,D}\hat{\Gamma}_{D,D}
  \hat{g}_{D,D}\\ &+&\hat{g}_{D,D}\hat{\Gamma}_{D,D}
  \hat{g}_{D,D}\hat{\Gamma}_{D,D} \hat{g}_{D,D}\hat{\Gamma}_{D,D}
  \hat{g}_{D,D} \hat{\Gamma}_{D,D} \hat{G}_{D,D}
  .
  \label{eq:Dyson-B}
\end{eqnarray}
We next group the nonlocal Green's functions pairwise,
as indicated by the notation $\langle\langle ... \rangle\rangle$,
and find

%%%%%%%%%%%%%%%%%%%%%%%%%%%
\begin{equation}
\label{eq:Dyson1-quatro}
\begin{aligned}
\langle\langle \hat{G}_{D,D} \rangle\rangle^{(\text{quartets})}
&=
\Big[
\hat{I}
-
\langle\langle
\hat{g}_{D,D}\hat{\Gamma}_{D,D}
\hat{g}_{D,D}\hat{\Gamma}_{D,D}
\hat{g}_{D,D}\hat{\Gamma}_{D,D}
\hat{g}_{D,D}\hat{\Gamma}_{D,D}
\rangle\rangle
\Big]^{-1}
\\
&\quad \times
\Big[
\langle\langle \hat{g}_{D,D} \rangle\rangle
+
\langle\langle
\hat{g}_{D,D}\hat{\Gamma}_{D,D}\hat{g}_{D,D}
\rangle\rangle
\\
&\qquad +
\langle\langle
\hat{g}_{D,D}\hat{\Gamma}_{D,D}
\hat{g}_{D,D}\hat{\Gamma}_{D,D}
\hat{g}_{D,D}
\rangle\rangle
\\
&\qquad +
\langle\langle
\hat{g}_{D,D}\hat{\Gamma}_{D,D}
\hat{g}_{D,D}\hat{\Gamma}_{D,D}
\hat{g}_{D,D}\hat{\Gamma}_{D,D}
\hat{g}_{D,D}
\rangle\rangle
\Big].
\end{aligned}
\end{equation}

%%%%%%%%%%%%%%%%%%
We then phenomenologically decompose $L=\langle\langle
\hat{g}_{D,D}\hat{\Gamma}_{D,D} \hat{g}_{D,D}\hat{\Gamma}_{D,D}
\hat{g}_{D,D}\hat{\Gamma}_{D,D} \hat{g}_{D,D}
\hat{\Gamma}_{D,D}\rangle\rangle$ into two terms $q$ and $L_1$. The former
couples to the Cooper-quartet phase, while the latter 
do not:
\begin{eqnarray}
  \label{eq:q}
  q&=& \langle\langle
  \hat{g}_{\alpha,\beta}^{1,1}\hat{\Gamma}_{\beta,\beta}^{1,2}
  \hat{g}_{\beta,\gamma}^{2,2}\hat{\Gamma}_{\gamma,\gamma}^{2,1}
  \hat{g}_{\gamma,\beta}^{1,1} \hat{\Gamma}_{\beta,\beta}^{1,2}
  \hat{g}_{\beta,\alpha}^{2,2}
  \hat{\Gamma}_{\alpha,\alpha}\rangle\rangle\\ L_1&=& \mbox{Sum of all
    remaining contractions}, \label{eq:remaining}
\end{eqnarray}
where the tight-binding sites $\alpha$, $\beta$, and $\gamma$ are the
normal-metal counterparts of $a$, $b$, and $c$ on the superconducting
sides of the interfaces between $N$ and $S_a$, $S_b$ and $S_c$. The
symbols $a$, $b$, and $c$ are used as a generic labels for the
superconductors and take values corresponding to all
permutations of $\{S_L,\,S_R,\,S_B\}$.
  %%%%%%%%%%%%%%%%%%%%%%%%%%%%%%%%%%%%%%%%%%%%%%%%%%%%%%%%%%%%%%%%%%%%%%%
\begin{figure}

  % Row 1: panels a and b
  \includegraphics[width=7 cm]{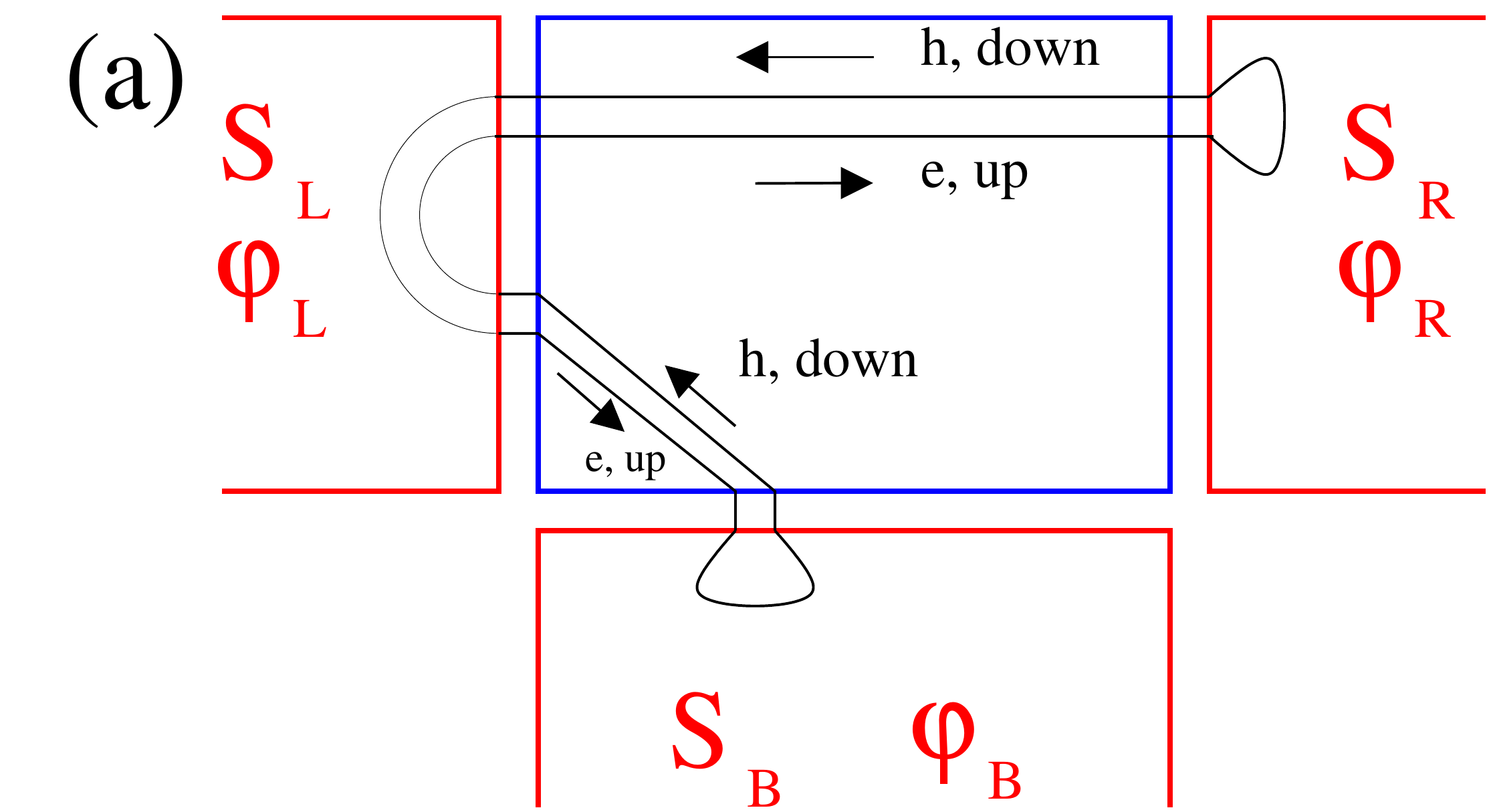}\hspace{0.02\textwidth}
  \includegraphics[width=7cm]{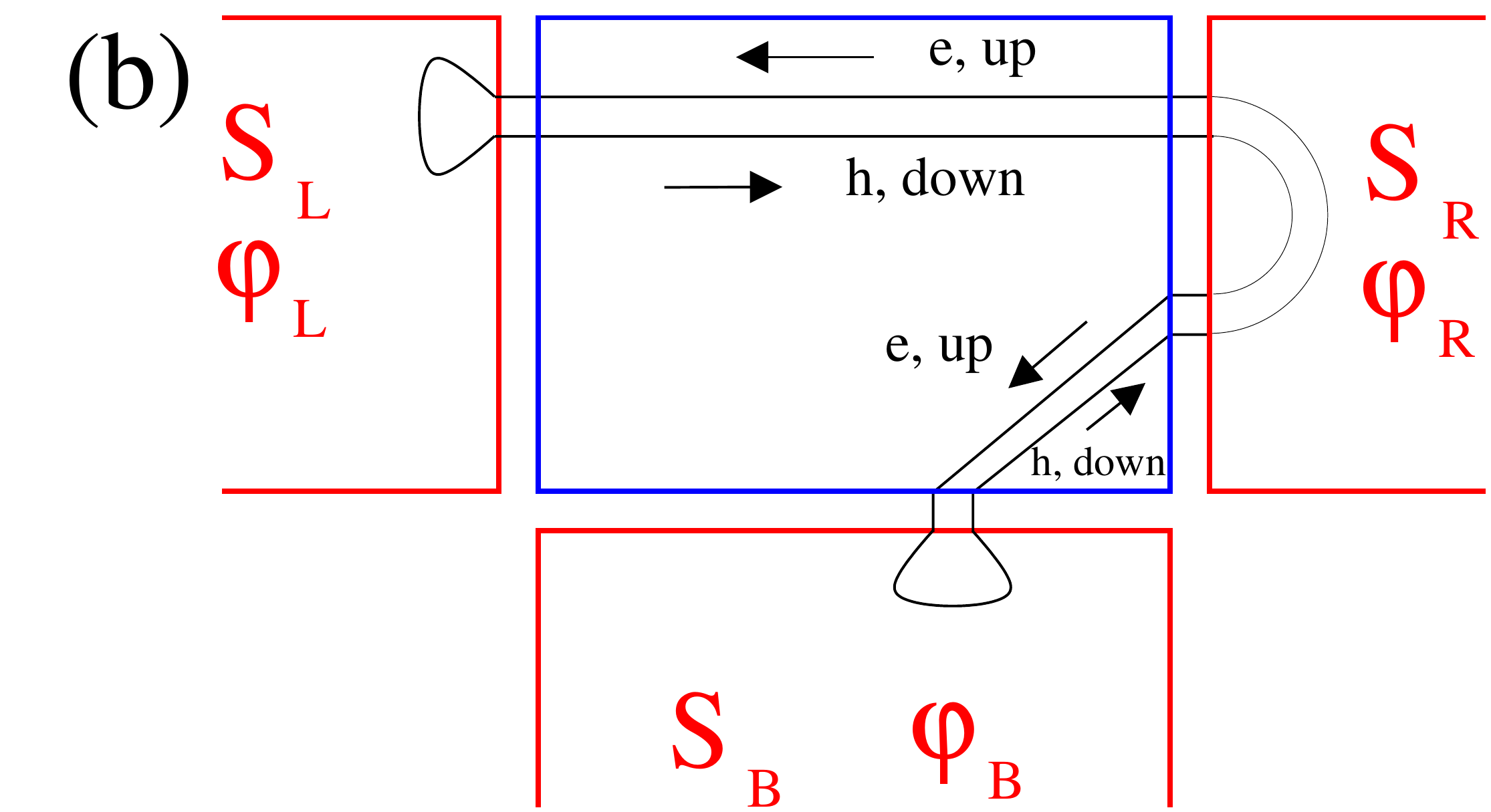}

  \vspace{0.8em}

  % Row 2: panel c (left aligned)
  \includegraphics[width=7 cm]{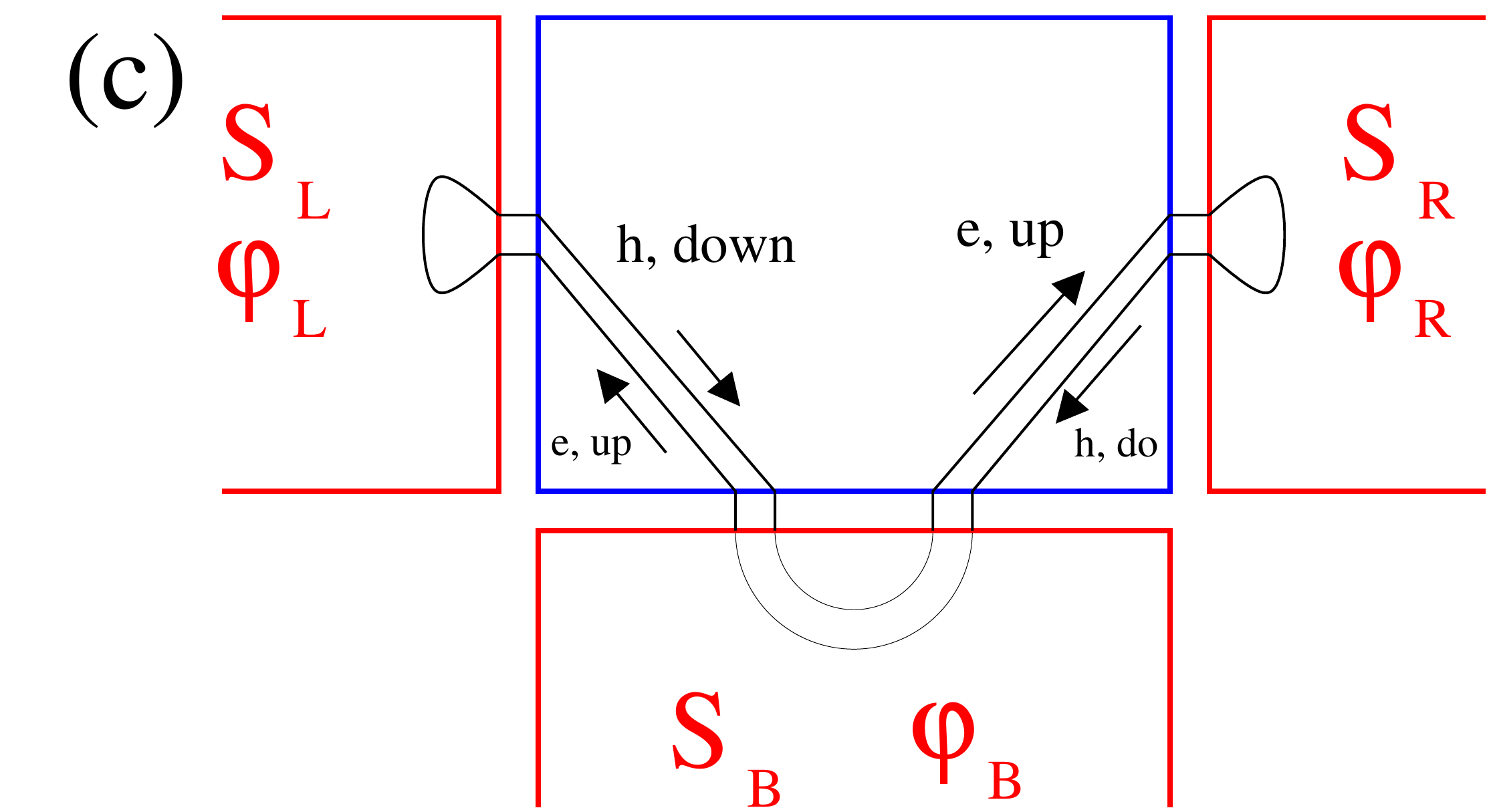}

  \caption{\textbf{Copper quartet diagrams for 3TJJ.}  (a-c) The three lowest-order Cooper quartet diagrams, corresponding
  to the $Q_L$ (a), $Q_R$ (b), and $Q_B$ (c) quartets. Those three types of Cooper quartets have the
  following Cooper quartet phase variables:
  $2\varphi_L-\varphi_R-\varphi_B$, $-\varphi_L+2\varphi_R-\varphi_B$
  and $-\varphi_L-\varphi_R+2\varphi_B$, respectively. The resonance lines in the experiment are interpreted as arising from $Q_L$ and
  $Q_R$ in panels (a) and (b), respectively.}
  \label{fig:Cooper-quartets}
\end{figure}
%%%%%%%%%%%%%%%%%%%%%%%%%%%%%%%%%%%%%%%%%%%%%%%%%%%%%%%%%%%%%%%%%%%%%%%

Figure~\ref{fig:Cooper-quartets} schematically shows the lowest order
Cooper quartet diagrams contributing to $q$; see Eq.~(\ref{eq:q}).  Panels (a), (b), and (c) of Fig.~\ref{fig:Cooper-quartets} show the three
types of Cooper quartets that appear at lowest order in tunneling:
the $Q_L$, $Q_R$ and $Q_B$ quartets, which transfer a charge $4e$
into $S_L$, $S_R$, and $S_B$, respectively.

The remaining terms $L_1$ in Eq.~(\ref{eq:remaining}) include contributions that are
insensitive to the superconducting phase variables. Other
contributions within $L_1$ depend on second harmonics of the
``local'' Josephson relations, such as $2(\varphi_L-\varphi_R)$. We find
\begin{equation}
  \label{eq:double-series}
  \frac{1}{1-q-L_1}= \sum_{n_1,n_2}
  \frac{(n_1+n_2)!}{n_1!n_2!}  q^{n_1} L_1^{n_2},
\end{equation}
where $n_1$ and $n_2$ two integers. We impose the condition of full
constructive interference by summing the simplest subseries with
$n_2=0$. From this, we deduce the following Cooper-quartet resonance energies:
\begin{equation}
  \label{eq:E-quartets0}
  E_n^{(quartets)}=\frac{\hbar v_F}{2(R_{\alpha,\beta}+R_{\beta,\gamma})}
  \left[2\pi n \mp \chi_{q}\right]
  ,
\end{equation}
where $\chi_{q} = \varphi_a + \varphi_b - 2
\varphi_c$ is one of the Cooper-quartet phase variables associated with
the superconductors $S_a$, $S_b$, and $S_c$. The
superconducting leads $S_a$, $S_b$, and $S_c$ with phases
$\varphi_a$, $\varphi_b$ and $\varphi_c$ can take all
permutations of the set $\{S_L,S_R,S_B\}$.

The Cooper-quartet spectrum in Eq.~(\ref{eq:E-quartets0}) is not
shifted by a half-period because the four constituent Andreev
reflections produce, at zero energy, a total phase shift of
$4\times (\pi/2)=2\pi$, which is equivalent to zero modulo $2\pi$.

We introduce the general phase shift $4\theta_{AR}$ associated
with the four Andreev reflections, leading to the following
phenomenological expression for the Cooper-quartet spectrum:
\begin{equation}
  \label{eq:E-quartets}
  E_n^{(quartets)}=\frac{\hbar v_F}{2(R_{\alpha,\beta}+R_{\beta,\gamma})}
  \left[2\pi n + 4 \theta_{AR} \mp \chi_{q}\right]
  .
\end{equation}

\section{Tunneling spectroscopy}
\label{sec:tunnel}

%%%%%%%%%%%%%%%%%%%%%%%%%%%%%%%%%%%%%%%%%%%%%%%%%%%%%%%%%%%%%%%%%%%%%%%
\begin{figure*}[htb]
  \includegraphics[width=15cm]{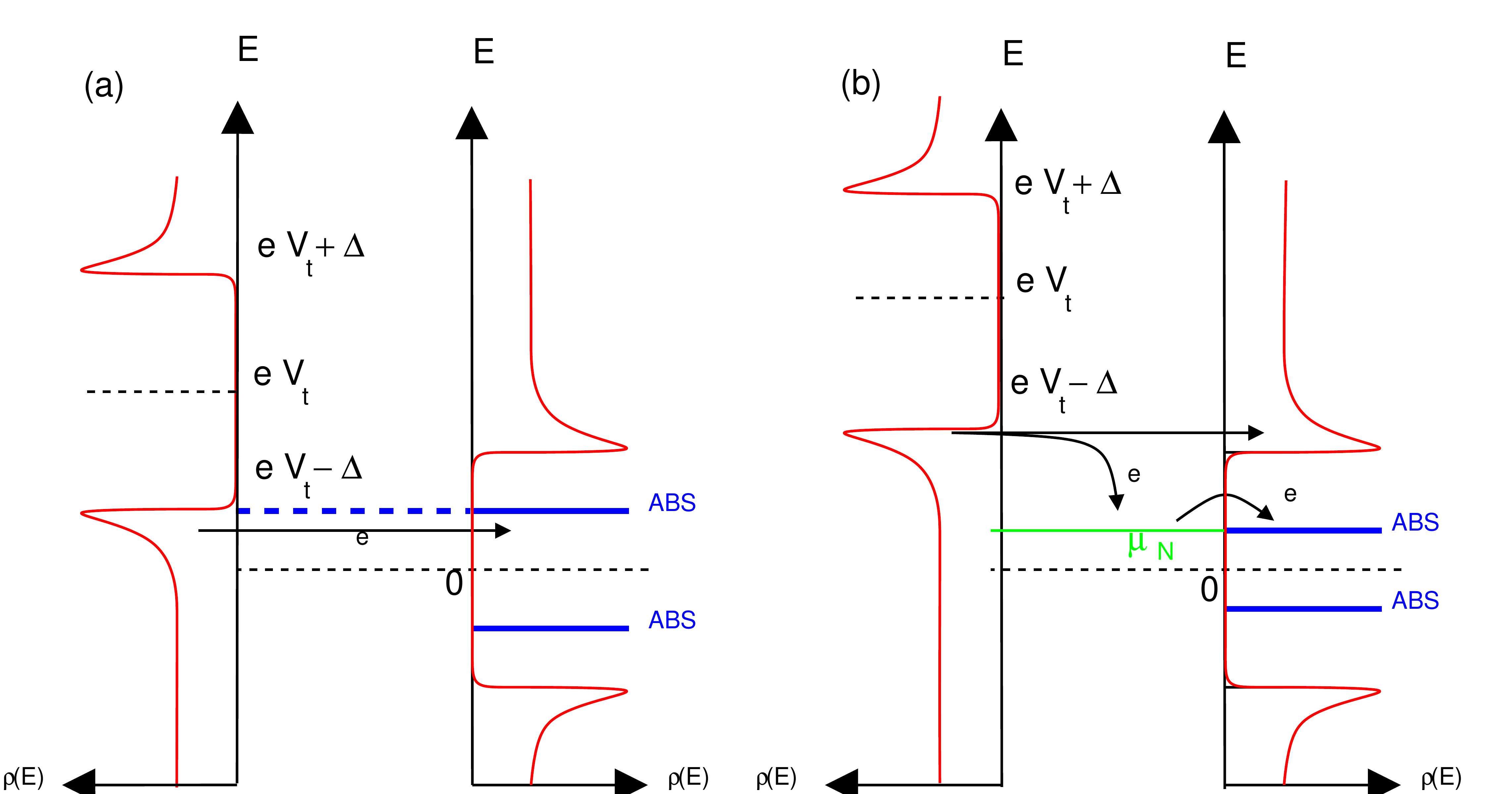}
      \caption{\textbf{Energy as a function of density of states.} The two spectroscopic modes: $V_t \leq 2\Delta$ (a) and $V_t \geq 2\Delta$ (b). The tunnel probe $S_t$ is biased at
        the voltage $V_t$. Density of states are shown for the left, right and bottom grounded
        superconducting leads $\{S_L,\,S_R,\,S_B\}$ in each panel. 
\label{fig:spectroscopic-modes}
}
\end{figure*}
%%%%%%%%%%%%%%%%%%%%%%%%%%%%%%%%%%%%%%%%%%%%%%%%%%%%%%%%%%%%%%%%%%%%%%%

Here, we discuss how superconducting tunneling
spectroscopy is implemented by a superconducting tunneling probe
$S_t$, biased at voltage $V_t$, to the 2D metal, in the presence of three 
superconducting contacts $\{S_L,\,S_R,\,S_B\}$
forming a three-terminal Josephson junction. Using the Keldysh formalism, we recalculate the
conductance in the Giaver regime \cite{Gaviour1960}, and establish a direct
connection with our experimental observation.

We first consider a weak hopping amplitude $\Sigma_{a,\alpha} =
\Sigma_{\alpha,a}\equiv \Sigma_a$ between the two superconductors
$S_a$ and $S_\alpha$, and evaluate the current to the order
$(\Sigma_a)^2$ in the presence of a bias voltage $V$. The ``1,1''
electron-electron Nambu component of the spectral current is given by
\begin{eqnarray}
  \label{eq:I-11}
  I_{1,1}(\omega)=\Sigma_{a,\alpha}^{1,1} \left(\hat{g}_{\alpha,\alpha}
  \Sigma_{\alpha,a} \hat{g}_{a,a}\right)^{+,-,1,1}(\omega,\omega)
  - \Sigma_{\alpha,a}^{1,1} \left(\hat{g}_{a,a} \hat{\Sigma}_{a,\alpha}
  \hat{g}_{\alpha,\alpha}\right)^{+,-,1,1}(\omega,\omega)
  ,
\end{eqnarray}
where the superscripts ``$+,-$'' denote the Keldysh Green's functions
\cite{Caroli1971,Caroli1972,Cuevas1}. In Eq.~(\ref{eq:I-11}), the spectral
current is diagonal both in the ``1,1'' electron-electron Nambu indices
and in the frequency $\omega$. We find
\begin{eqnarray}
  I_{1,1}(\omega)&=&
  \hat{\Sigma}_{a,\alpha}^{1,1} \hat{g}^{+,-,1,1}_{\alpha,\alpha}(\omega+eV)
  \hat{\Sigma}_{\alpha,a}^{1,1} \hat{g}^{A,1,1}_{a,a} (\omega)
  + \hat{\Sigma}_{a,\alpha}^{1,1} \hat{g}^{R,1,1}_{\alpha,\alpha}(\omega+eV)
  \hat{\Sigma}_{\alpha,a}^{1,1} \hat{g}^{+,-,1,1}_{a,a} (\omega)\\
  &&-
  \hat{\Sigma}_{\alpha,a}^{1,1} \hat{g}^{+,-,1,1}_{a,a}(\omega-eV)
  \hat{\Sigma}_{a,\alpha}^{1,1} \hat{g}^{A,1,1}_{\alpha,\alpha}(\omega)
  -
  \hat{\Sigma}_{\alpha,a}^{1,1} \hat{g}^{R,1,1}_{a,a}(\omega-eV)
  \hat{\Sigma}_{a,\alpha}^{1,1} \hat{g}^{+,-,1,1}_{\alpha,\alpha}(\omega)
  .
\end{eqnarray}
Additionally, taking into account the spectral current in the ``2,2''
hole-hole channel leads to the following expression for the current
as a function of the bias voltage $V$:
\begin{eqnarray}
  \label{eq:I-V}
  I(V)= \int I(\omega) d\omega = 8 \pi^2\left(\Sigma_a\right)^2 \int
  d\omega \rho_{\alpha,\alpha}(\omega) \rho_{a,a}(\omega-eV) \left[
    n_F(\omega-eV)-n_F(\omega)\right] ,
\end{eqnarray}
where the spectral current is $I(\omega)=I_{1,1}(\omega)
-I_{2,2}(\omega)$, and $\rho_{a,a}(\omega)$ and
$\rho_{\alpha,\alpha}(\omega)$ denote the spectral densities of
states of $S_a$ and $S_\alpha$, respectively. Reversing the signs of both
the bias voltage $V$ and the frequency $\omega$ in Eq.~(\ref{eq:I-V})
yields $I(-V)=-I(V)$. The local density of states $\rho(\omega)$ of
each BCS superconductor is proportional to the imaginary part of
the bare Green's function in the electron-electron channel:
$\rho(\omega)=\mbox{Im}[g^A(\omega)]/\pi$, where
$g^A(\omega)=-(\omega-i\eta)/W\sqrt{|\Delta|^2
-(\omega-i\eta)^2}$.

%%%%%%%%%%%%%%%%%%%%%%%%%%%%%%%%%%%%%%%%%%%%%%%%%%%%%%%%%%%%%%%%%%%%%%%
\begin{figure}[htb]
  \centering
    
      \includegraphics[width=10 cm]{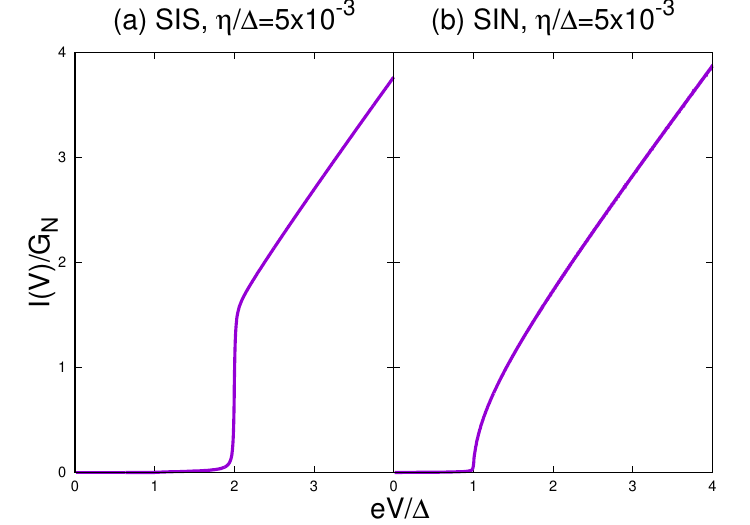}

      \caption{\textbf{The current-voltage characteristics.} (a-b) The current-voltage characteristics of a
      superconductor-superconductor weak link (a) and a
      superconductor-normal metal weak link (b). The voltage
      $V$ is normalized to the superconducting gap $\Delta$ and the current $I$ is normalized to the normal-state
      conductance $G_N$. The ratio between the Dynes
      parameter $\eta$ and the superconducting gap $\Delta$ is
      $\eta/\Delta=5\times 10^{-3}$ on each panel.}
      \label{fig:figure1}

\end{figure}
%%%%%%%%%%%%%%%%%%%%%%%%%%%%%%%%%%%%%%%%%%%%%%%%%%%%%%%%%%%%%%%%%%%%%%%

Figure~\ref{fig:figure1} shows the current-voltage characteristics
derived from Eq.~(\ref{eq:I-V}), for Giaver tunneling in both
a $SIS$ weak link [Fig.~\ref{fig:figure1}(a)] and a $SIN$ weak
link [see Fig.~\ref{fig:figure1}(b)].

We find that biasing a SIS tunnel junction
at voltages $|eV_t|<2 \Delta$ results in the absence of tunneling
current between the tunneling probe and any other superconducting terminal.

\section{Cooper quartet resonances in 2D}
\label{sec:2D}

We now focus specifically on ABS for a 2D normal metal, in contrast
to the 1D and 3D cases discussed earlier.
Solutions of the wave equation in even dimensions exhibit a {\it wake}
effect that is absent in odd dimension. We show that this wake effect
has important consequences for the dependence of the tunneling spectra
on the value of the bias voltage $V_t$ applied to the tunneling probe.

Specifically, we obtain the following expressions for the nonlocal
Green's function in 2D:
\begin{eqnarray}
  \label{eq:g2D-1}
  g_{\alpha,\beta,1,1}^{2D} \simeq \frac{i}{W\sqrt{k_F R}}
  \cos{\left[\left(k_F+\frac{\omega}{v_F}\right)
      R-\frac{\pi}{4}\right]}\\ g_{\alpha,\beta,2,2}^{2D} \simeq
  \frac{i}{W\sqrt{k_F R}}
  \cos{\left[\left(k_F-\frac{\omega}{v_F}\right)
      R-\frac{\pi}{4}\right]}
  .
  \label{eq:g2D-2}
\end{eqnarray}
These expressions contrast with the 1D and 3D counterparts given in
Eqs.~(\ref{eq:g-1D-1})-(\ref{eq:g-1D-2}) and
Eqs.~(\ref{eq:g-3D-1})-(\ref{eq:g-3D-2}), respectively.  We next
substitute Eqs.~(\ref{eq:g2D-1})-(\ref{eq:g2D-2}) into the kernel of
Eq.~(\ref{eq:kernel}) and note that the product
$g_{\alpha,\beta,1,1}^{2D} g_{\alpha,\beta,2,2}^{2D}$ is
real-valued. In 2D, the condition for full constructive interference therefore
becomes equivalent to fixing the phase: $\chi_{p} = \varphi_R -
\varphi_L = 2 \theta_{AR}+2\pi n$, where $n$ is an integer, and
$\theta_{AR}$ is defined in
Eq.~(\ref{eq:theta-pairs}). 

%%%%%%%%%%%%%%%%%%%%%%%%%%%%%%%%%%%%%%%%%%%%%%%%%%%%%%%%%%%%%%%%%%%%%%%
\begin{figure*}[htb]
  \centering
  
\includegraphics[width=10 cm]{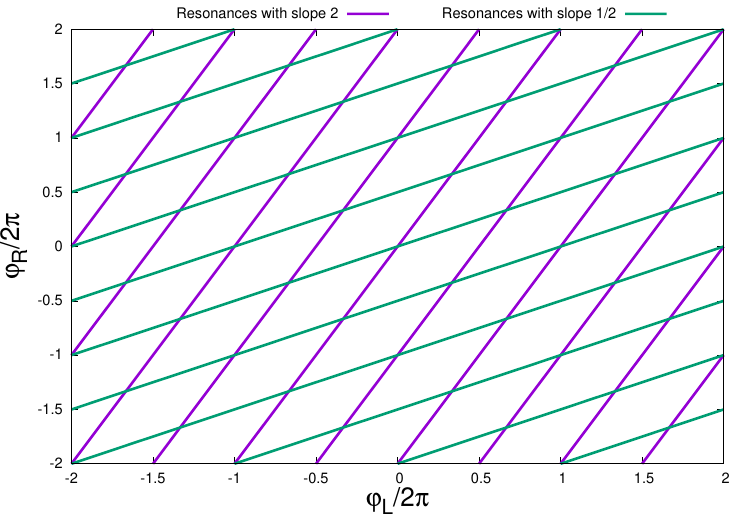}
\caption{\textbf{The two families of the Cooper quartet resonances:} The
  resonances with slope $2$ in the $(\varphi_L,\,\varphi_R)$ diagram,
  and the resonances with slope $1/2$, as defined in Eqs.~(\ref{eq:21})
  and~(\ref{eq:12}).
\label{fig:schema-resos}
}
\end{figure*}
%%%%%%%%%%%%%%%%%%%%%%%%%%%%%%%%%%%%%%%%%%%%%%%%%%%%%%%%%%%%%%%%%%%%%%%

A generalization to the Cooper quartets leads to $\chi_{q} =
4\theta_{AR} + 2\pi n$. The resulting
Cooper quartet resonance lines are given by:
\begin{eqnarray}
  \label{eq:21-0}
  2\varphi_L - \varphi_R &=& 4\theta_{AR} + 2 \pi n\\
  - \varphi_L + 2 \varphi_R &=& 4\theta_{AR}+2\pi m
  \label{eq:12-0}
  ,
\end{eqnarray}
These two relationships correspond to resonance lines with slopes $1/2$ and $2$ in the
$(\varphi_L,\,\varphi_R)$ plane of the superconducting phase
differences, respectively. Here, we set $\varphi_B=0$, as the phase reference. These equations can also be rearranged as follows:
\begin{eqnarray}
  \label{eq:21}
   \varphi_R = 2 \varphi_L - 4\theta_{AR}-2\pi n\\ \varphi_R
   =\frac{1}{2} \varphi_L + 2\theta_{AR} + \pi m
  \label{eq:12}
  .
\end{eqnarray}
We note that Eqs.~(\ref{eq:21})-(\ref{eq:12}) hold for all energies.

The above Eqs.~(\ref{eq:21})-(\ref{eq:12}) are schematically
illustrated in Fig.~\ref{fig:schema-resos}, highlighting the two families of Cooper-quartet resonances.
Within this simple model, the two families intersect freely without hybridization. 
By contrast, the experimental data reveal avoided crossings, indicating hybridization
between the corresponding quartet modes.

\section{Theoretical Cooper-quartet
  diagram}

\label{sec:fit-manips}

We begin by evaluating the Cooper-quartet diagram shown in
Fig.~\ref{fig:Cooper-quartets} (a). The spectral current transmitted from
the 2D metal $N$ to the superconductor $S_R$ takes the following form:
\begin{equation}
  I_R(\omega)=\mbox{Nambu-trace}\left\{\hat{\sigma}_z
  \left[ \hat{\Sigma}_{R,r} \hat{G}^{+,-}_{r,R}
    -\hat{\Sigma}_{r,R} \hat{G}^{+,-}_{R,r} \right]\right\}
  .
\end{equation}
Expanding this equation, we find
\cite{recent-preprint}
\begin{eqnarray}
  \label{eq:dIRdmiN}
  \frac{\partial I_R}{\partial \mu_N} &\approx& \frac{\gamma_R
    \gamma_L^2 \gamma_B}{W^4} \frac{1}{(k_F R_{r,l})(k_F R_{l,b})}
  \cos\left(\frac{2\mu_N R_{r,l}}{\hbar v_F}\right)
  \cos\left(\frac{2\mu_N R_{l,b}}{\hbar v_F}\right) \cos\left(2
    \varphi_L - \varphi_R - \varphi_B\right) .
\end{eqnarray}
Here, we have discarded a prefactor of order unity and chosen the
superconducting phase variables such that Eq.~(\ref{eq:21}) is
satisfied, i.e., $2\varphi_L-\varphi_R-\varphi_B = 4\theta_{AR} + 2\pi
n$ in Fig.~\ref{fig:Cooper-quartets} (a).  The
labels $R$, $L$, and $B$ stand for the tight-binding sites supporting
the Cooper quartet resonances in the superconducting leads, while
their counterparts in the normal metal $N$ are denoted by $r$, $l$,
and $b$, respectively. We also note that current conservation relates
the tunneling conductance to the Cooper-quartet current: the condition
$I_t+I_R+I_L+I_B=0$ implies $\partial I_t/\partial \mu_N= -\partial
(I_R+I_L+I_B)/\partial \mu_N$.

We now assume that the dominant contribution to the current arises
from the $(l,b)$ section of the quartet diagram that is located near the
bottom-left corner of the normal-metal
$N$; see Fig.~\ref{fig:Cooper-quartets} (a). This assumption maximizes the 
 product of the geometrical prefactors $1/{(k_F R_{r,l}) (k_F
  R_{l,b})}$. Therefore, we find
\begin{eqnarray}
- \frac{\partial I_R}{\partial \mu_N} &\approx& - \frac{\gamma_R
  \gamma_L^2 \gamma_B}{W^4} \frac{1}{(k_F R_{r,l})}
\cos\left(\frac{2\mu_N R_{r,l}}{\hbar v_F}\right) \cos\left(2
\varphi_L - \varphi_R - \varphi_B\right) ,
\end{eqnarray}
where we implemented $R_{l,b}=0$ in the
  $\cos(2\mu_N R_{l,b}/ \hbar v_F)$ term of Eq.~(\ref{eq:dIRdmiN})
  and, taking the local limit of the $(l,b)$ contact in the
  geometrical prefactor, we replaced $k_F R_{l,b}$ by unity.
 We conclude the
following phenomenological expression of the dimensionless
conductance:
\begin{equation}
  \label{eq:g0}
  g_0\left(\varphi_L,\varphi_R,V_{t,eff}\right)
  = - \cos\left(\alpha \mu^*_{N,eff}(V_{t,eff})\right)
  \delta_{eff} \left(2\varphi_L-\varphi_R-4\theta_{AR}\right)
  - \cos\left(\alpha \mu^*_{N,eff}(V_{t,eff})\right)
  \delta_{eff} \left(2\varphi_R-\varphi_L-4\theta_{AR}\right)
  ,
\end{equation}
where we have included both types of Cooper-quartet resonances (Eqs.~(\ref{eq:21})-(\ref{eq:12})) and phenomenologically introduced the nonequilibrium
electrochemical potential
$\mu_{N,eff}^*(V_{t,eff})=\sqrt{V_{t,eff}^2-(2\Delta)^2}$. The notation $\delta_{eff}$ in
Eq.~(\ref{eq:g0}) denotes a $\delta$-function broadened by a
Lorentzian along both the $\varphi_R$- and the $\varphi_L$-axes:
\begin{eqnarray}
  \label{eq:g0-2}
  g_0\left(\varphi_L,\varphi_R,V_{t,eff}\right)
  &=& - \cos\left(\alpha \mu^*_{N,eff}(V_{t,eff})\right) \int \frac{dt}{\pi^2}
  \frac{\eta_0}{\left[\left(\varphi_R-\varphi_{R,1}(t)\right)^2 + \eta_0^2\right]}
  \frac{\eta_0}{\left[\left(\varphi_L-\varphi_{L,1}(t)\right)^2 + \eta_0^2\right]}\\
  &&- \cos\left(\alpha \mu^*_{N,eff}(V_{t,eff})\right) \int \frac{dt}{\pi^2}
  \frac{\eta_0}{\left[\left(\varphi_R-\varphi_{R,2}(t)\right)^2 + \eta_0^2\right]}
  \frac{\eta_0}{\left[\left(\varphi_L-\varphi_{L,2}(t)\right)^2 + \eta_0^2\right]}
  ,
\end{eqnarray}

%%%%%%% I Read Up to Here %%%%%%

where $2\varphi_{L,1}(t) - \varphi_{R,1}(t) = 4\theta_{AR}$ and
$\varphi_{L,2}(t) - 2\varphi_{R,2}(t) = 4\theta_{AR}$ for both Cooper
quartet resonance lines in Eqs.~(\ref{eq:12}) and~(\ref{eq:21}),
respectively, i.e.
\begin{eqnarray}
  \varphi_{L,1}(t)=\frac{t}{2}\\
  \varphi_{R,1}(t)=t+4\theta_{AR}
\end{eqnarray}
and
\begin{eqnarray}
  \varphi_{L,2}(t)=t+4\theta_{AR}\\
  \varphi_{R,2}(t)=\frac{t}{2}
  .
\end{eqnarray}
Fig.~\ref{fig:figure_Vt_varie} shows the resulting conductance maps
for a selection of the dimensionless voltages $V_{t,eff}$. We conclude
that the model is qualitatively compatible with the main features of
the experiment.

%%%%%%%%%%%%%%%%%%%%%%%%%%%%%%%%%%%%%%%%%%%%%%%%%%%%%%%%%%%%%%%%%%%%%%%
\begin{figure*}[!t]
\centering
\includegraphics[width=\textwidth]{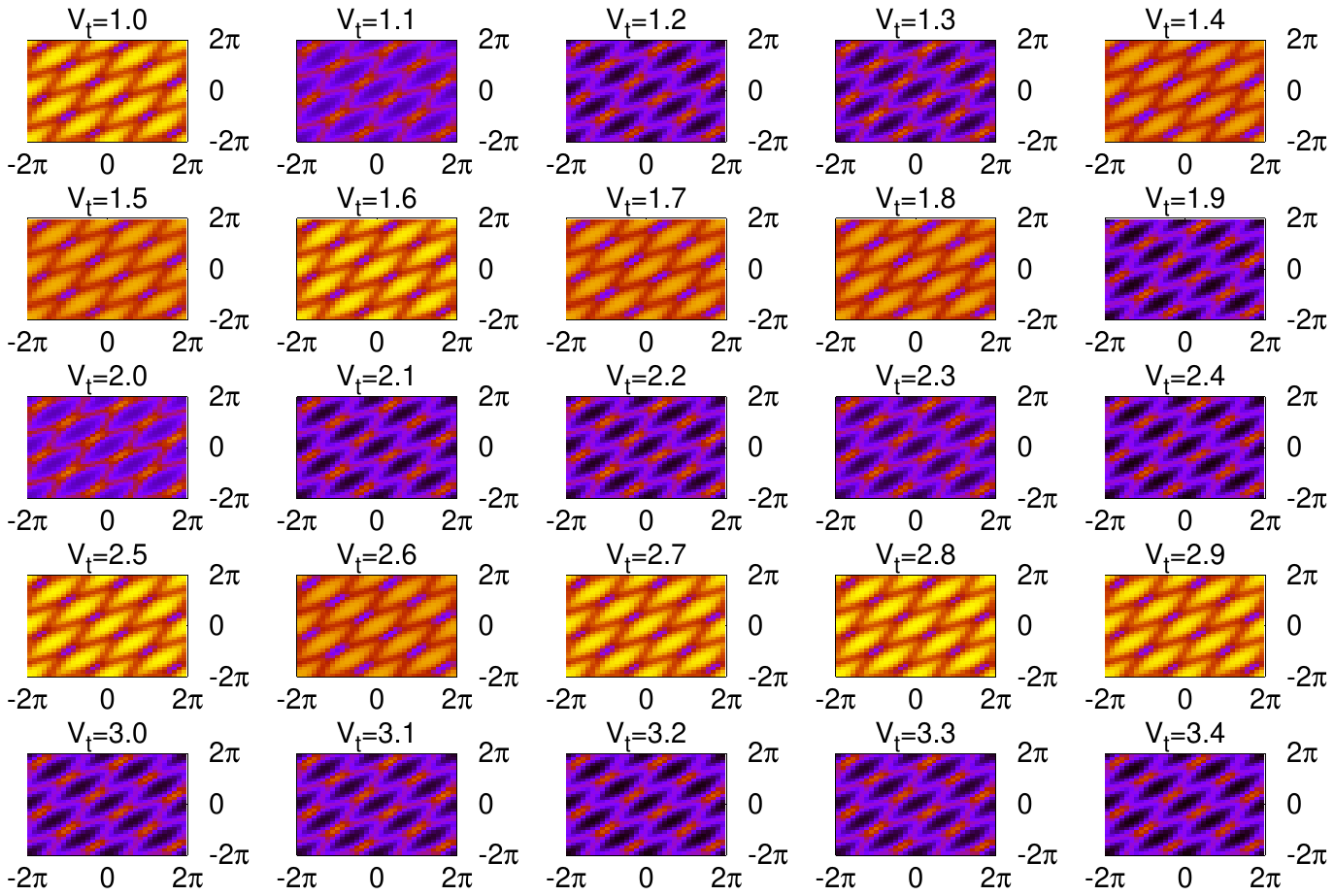}
\caption{\textbf{Theoretically calculated tunneling conductance maps.} Calculated tunneling conductance in the plane of the two
  phase variables $(\varphi_L,\varphi_R)$, for the different values of
  dimensionless $V_{t,eff}$.  The value
  $V_{t,eff}=1$ corresponds to bias voltage equal to twice the
  superconducting gap. We find evidence for the two families of
  resonances. As a result of the 2D quantum wake effect, the location
  of the resonances does not move as a function of $V_{t,eff}$ in the
  $(\varphi_L,\,\varphi_R)$ plane, only their sign oscillates between
  negative and positive values.
\label{fig:figure_Vt_varie}
}
\end{figure*}
%%%%%%%%%%%%%%%%%%%%%%%%%%%%%%%%%%%%%%%%%%%%%%%%%%%%%%%%%%%%%%%%%%%%%%%
%%%%%%%%%%%%%%%%%%%%%%%%%%%%%%%%%%%%%%%%%%%%%%%%%%%%%%%%%%%%%%%%%%%%%%%
\begin{figure*}[htb]
  \centering

  \begin{minipage}[t]{0.48\textwidth}
    \includegraphics[width=8 cm]{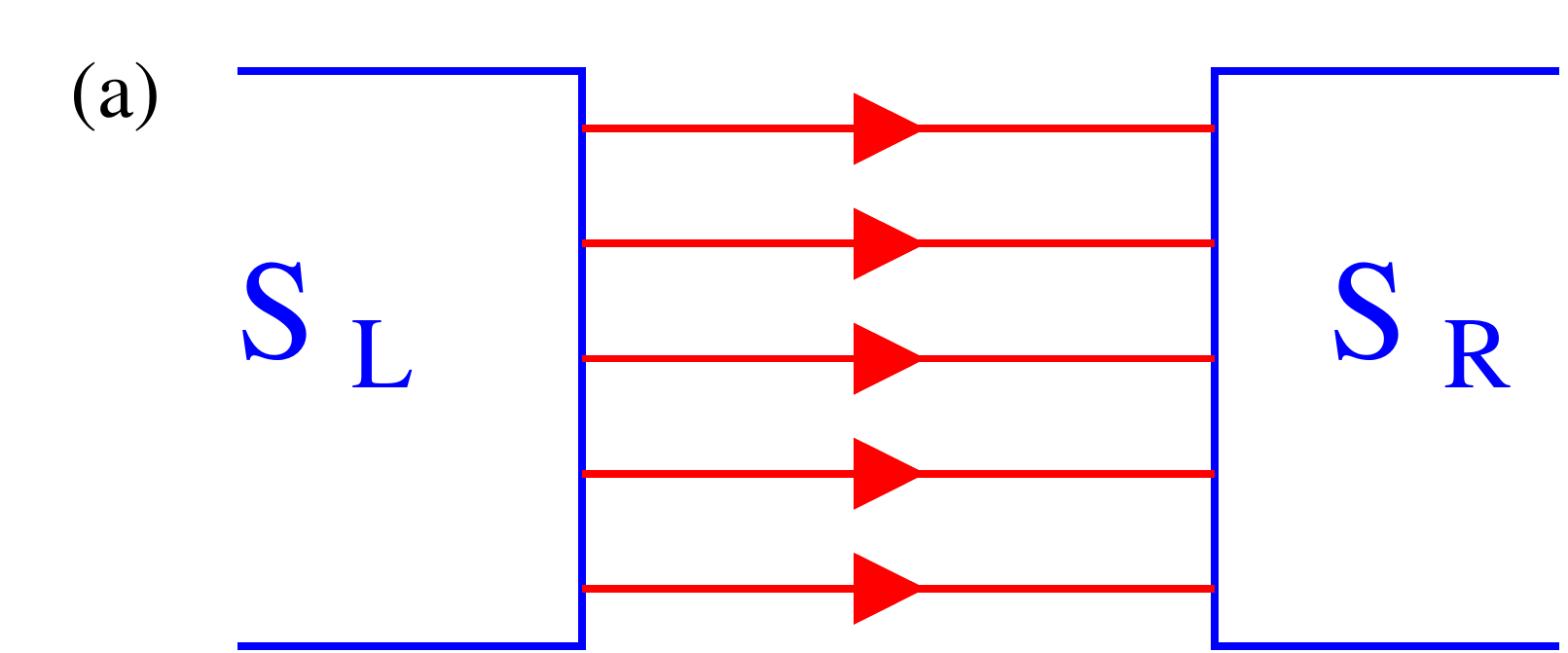}
  \end{minipage}\hfill
  \begin{minipage}[t]{0.48\textwidth}
    \includegraphics[width=8cm]{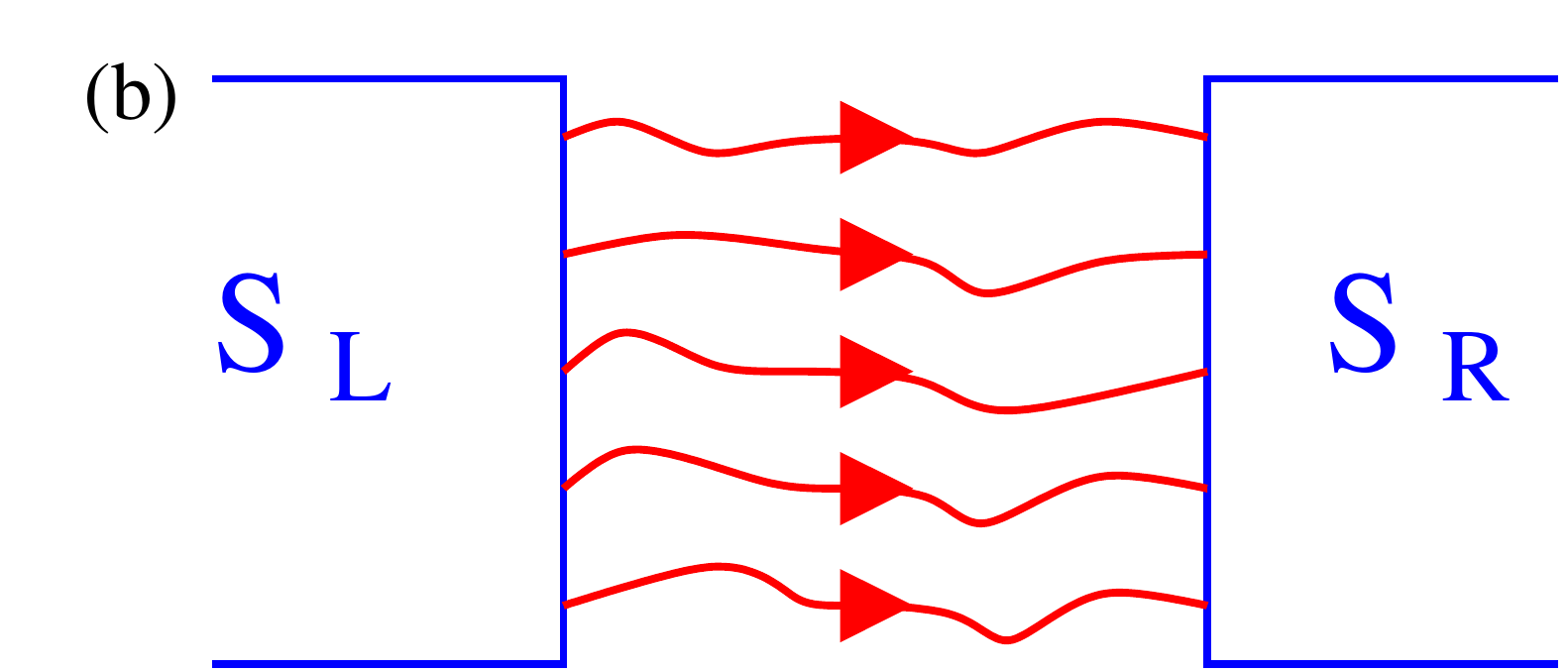}
  \end{minipage}

  \vspace{1em}

  \begin{minipage}[t]{0.48\textwidth}
    \includegraphics[width=8cm]{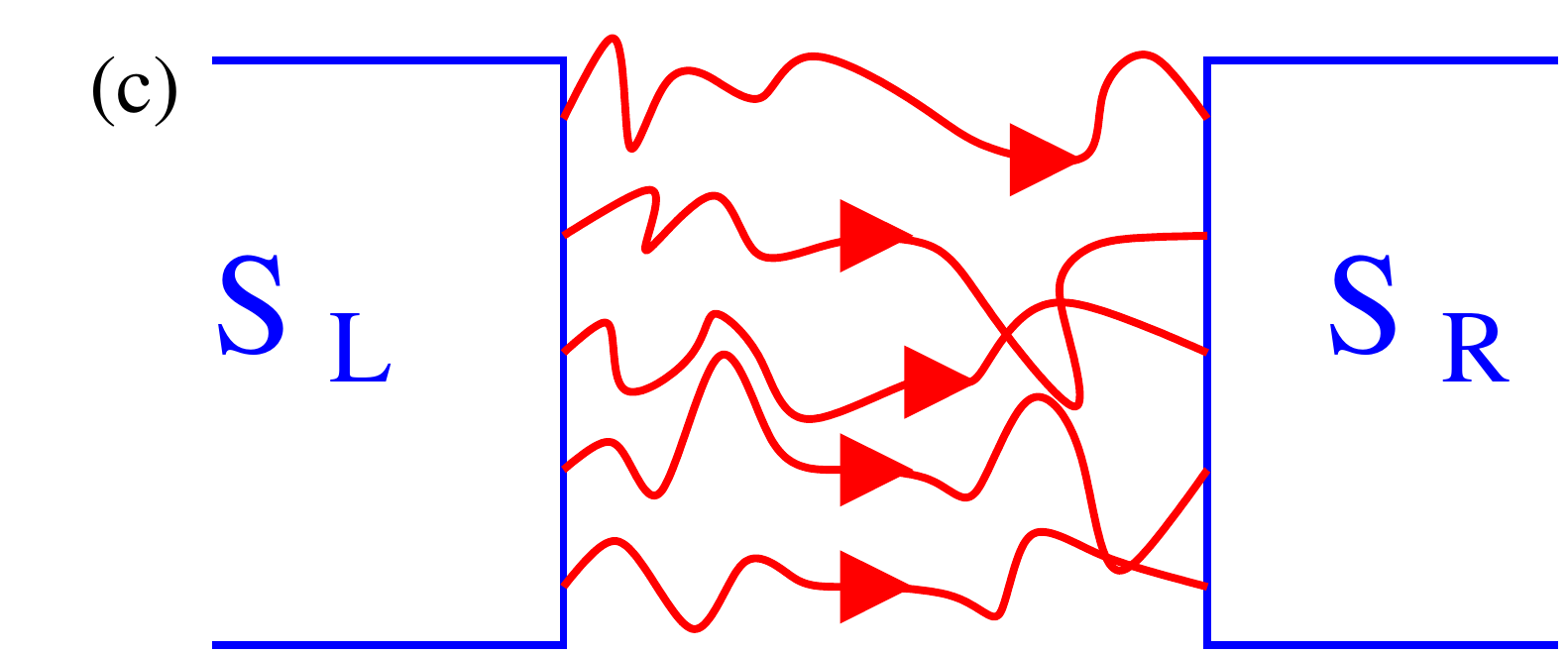}
  \end{minipage}

  \caption{\textbf{Andreev tubes.} Typical shape of the Andreev tubes: (a) in the absence of
  disorder, (b) with weak disorder, and (c) in the diffusive limit.}
  \label{fig:disorder}
\end{figure*}
%%%%%%%%%%%%%%%%%%%%%%%%%%%%%%%%%%%%%%%%%%%%%%%%%%%%%%%%%%%%%%%%%%%%%%%
%\end{figure}
\section{Robustness of the Cooper quartet resonances}

\subsubsection{Robustness with respect to weak disorder}

In this subsubsection, we argue that the Cooper quartet resonances are
immune with respect to weak disorder. Coming back to a long
two-terminal $SNS$ Josephson junction, we note that different
situations can be met upon increasing the amount of
disorder. Fig.~\ref{fig:disorder} (a) shows the Andreev tubes in a clean
$SNS$ device. Those tubes are parallel to the $x$-axis, due to the
conserved quantum numbers of the transverse modes forming standing
wave in the considered perfect wave-guide geometry. The experimental
devices of the paper are as clean as possible, due to the
encapsulation of the graphene layer by two hBN layers on top and
bottom, and to the contacts at the edges. Small
  residual disorder can distort the Andreev tubes as shown in
  Fig.~\ref{fig:disorder} (b). On this figure, the elastic mean free path
  is large and summing over all of the transverse channels yields the
  oscillating factors in the transmission modes deduced from
  Eqs.~(\ref{eq:g2D-1})-(\ref{eq:g2D-2}):
\begin{eqnarray}
  T&\sim& \frac{1}{\delta l} \int_{l_0-\delta l/2}^{l_0+\delta l/2} dl
  \cos{\left[\left(k_F+\frac{\omega}{v_F}\right)
      l-\frac{\pi}{4}\right]}
  \cos{\left[\left(k_F-\frac{\omega}{v_F}\right)
      l-\frac{\pi}{4}\right]} .
\end{eqnarray}
It is deduced that all Andreev tubes keep their coherence upon
including a distribution ${\cal P}(l)$ of the tube length $l$,
characterized by the root-mean-square $\delta l$ such that $2\omega
\delta l / \hbar v_F\lesssim 1$, see Fig.~\ref{fig:disorder} (b).  As the
level of disorder is increased, the typical
semiclassical trajectories start to scatter between the nonmagnetic
impurities, see Fig.~\ref{fig:disorder} (c). The semiclassical
trajectories connecting the left to the right side of the normal metal
$N$ are characterized by the root-mean-square $\delta l\gg\hbar
v_F/2\omega$ in their distribution. As a result of the corresponding
destructive interference, the energy oscillations of the supercurrent
are damped by a strongly decaying exponential envelope above the
Thouless energy
\cite{Nakano,Zaitsev,Kadigrobov,Volkov,Nazarov1,Nazarov2,Wilhelm,Yip,Belzig,Zaikin1,Zaikin2,Zaikin3,Zaikin4}. This
contrasts with the coherent oscillations in the clean limit
\cite{Melin3TJJ,NonequilibriumABS2025}.

%%%%%%%%%%%%%%%%%%%%%%%%%%%%%%%%%%%%%%%%%%%%%%%%%%%%%%%%%%%%%%%%%%%%%%%
\begin{figure*}
  \begin{flushleft}
  \begin{minipage}{.4\textwidth}
    \includegraphics[width=\textwidth]{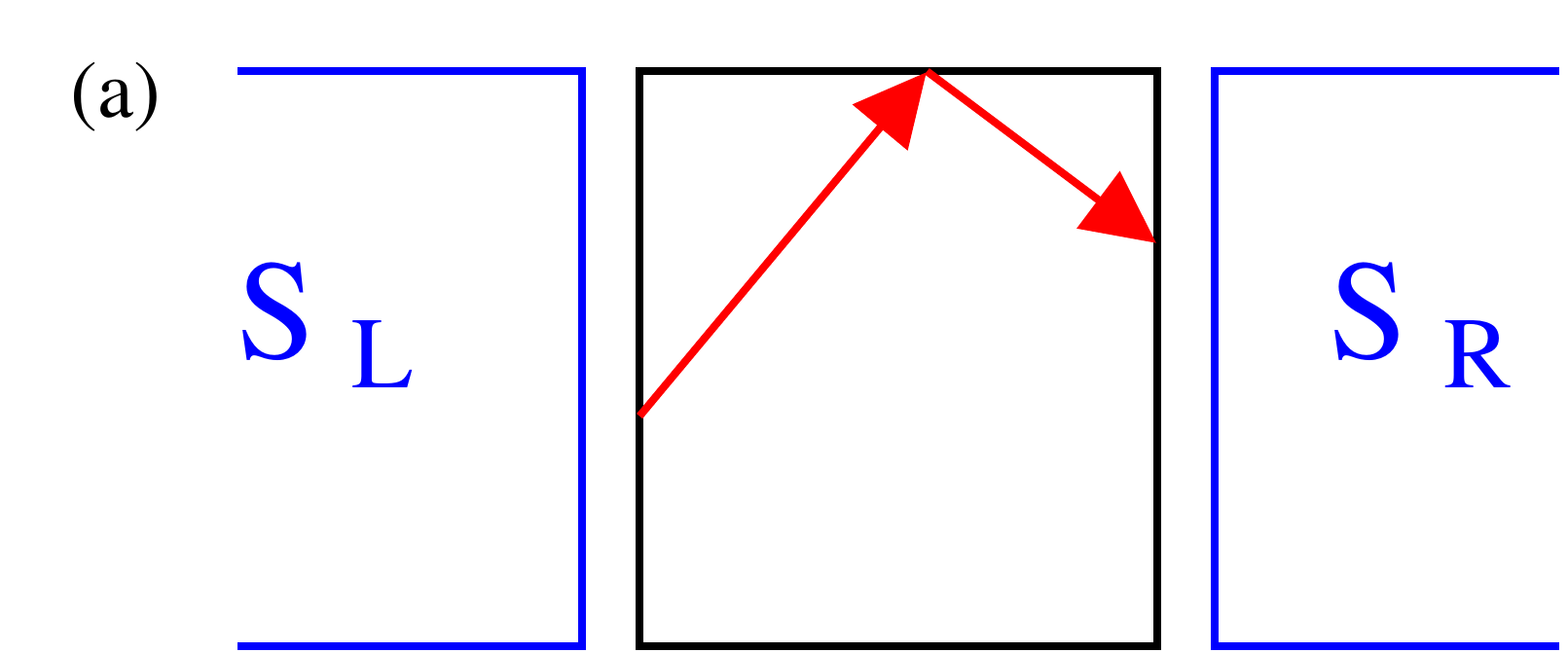}
  \end{minipage}\end{flushleft}\vspace*{-4.2cm}\begin{flushright}
      \begin{minipage}{.4\textwidth}
    \includegraphics[width=\textwidth]{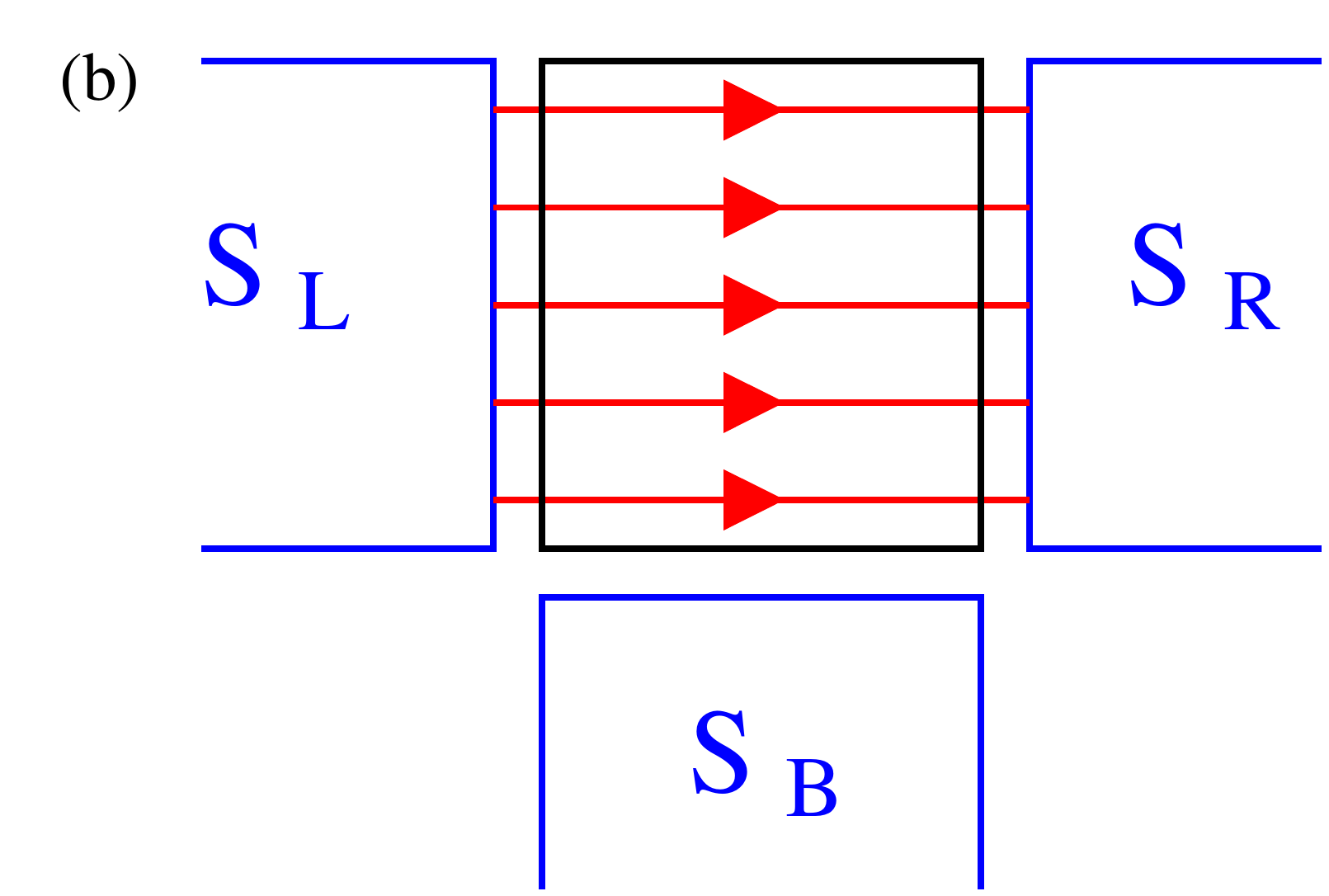}
  \end{minipage}\end{flushright}

  \begin{flushleft}
  \begin{minipage}{.4\textwidth}
    \includegraphics[width=\textwidth]{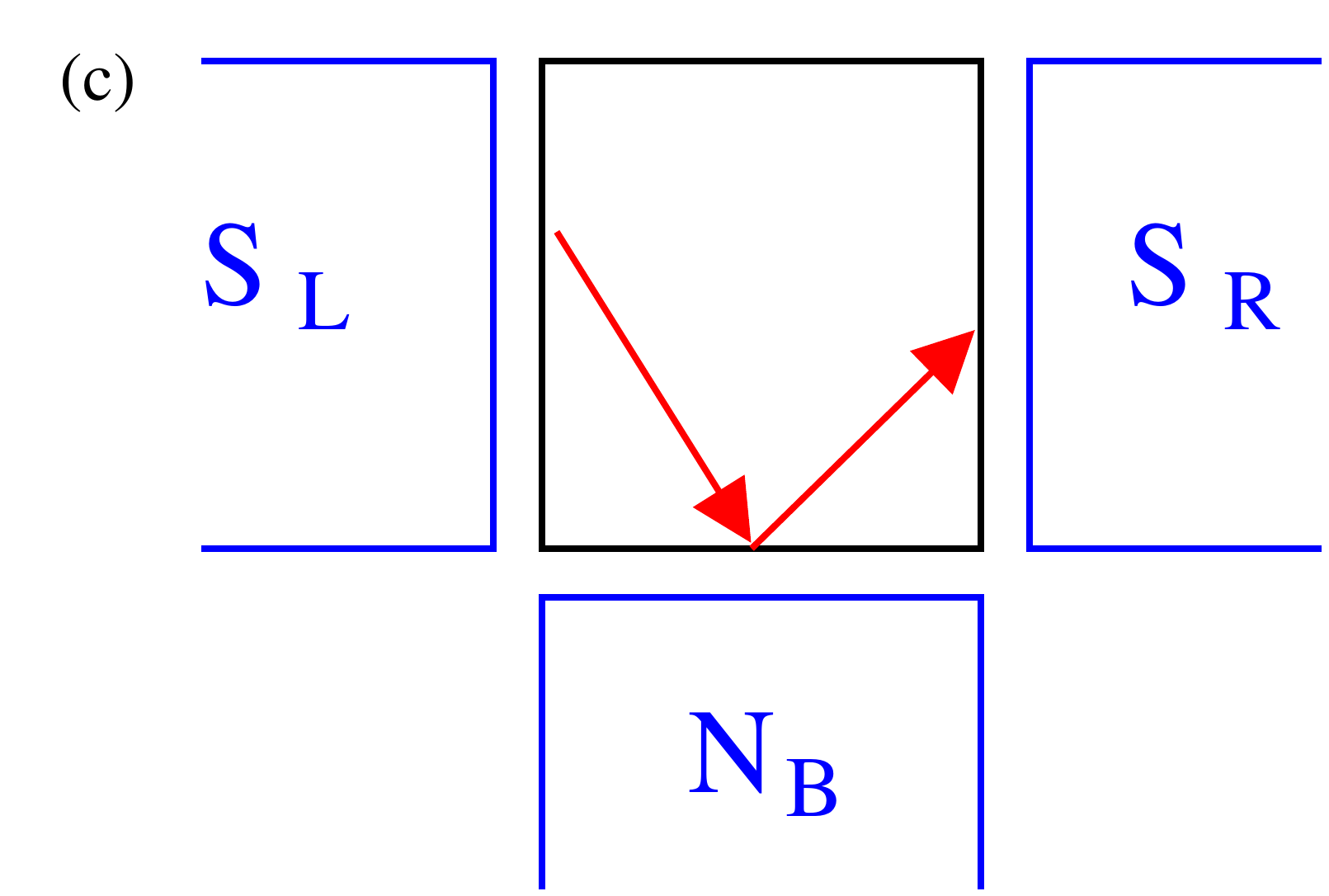}
  \end{minipage}\end{flushleft}
    \caption{\textbf{Reflection of the Andreev tubes at an open boundary.} (a),
      absence of reflection at the boundary with the superconductor
      $S_B$ (b) and reflection at the boundary with the normal metal
      $N_B$ (c).
\label{fig:AT}
}
\end{figure*}
%%%%%%%%%%%%%%%%%%%%%%%%%%%%%%%%%%%%%%%%%%%%%%%%%%%%%%%%%%%%%%%%%%%%%%%

\subsubsection{Robustness with respect to scattering Cooper pairs
  at the 2D metal-$S$ boundaries}

We note that scattering Andreev tubes at the open boundary of the 2D
metal, see Fig.~\ref{fig:AT} (a), can change the magnetic field
dependence of the corresponding Fraunhofer pattern,
see Ref.~\cite{Meier2016}. In this
subsection, we demonstrate that scattering Andreev tubes at the 2D
metal-superconductor interfaces, see Fig.~\ref{fig:AT} (b), does not
contribute to the current. Conversely, scattering Andreev tubes at a
2D metal-normal metal interface, see Fig.~\ref{fig:AT} (c),
contributes for a finite value to the current.

We start with evaluating the current associated to Andreev tubes
propagating between the interfaces with the left and right
superconductors $S_L$ and $S_R$, and bouncing once on the interface
between the 2D metal and the bottom superconductor $S_B$, see the
diagram in Fig.~\ref{fig:AT} (b). The current $I_L$ transmitted at the
left contact with $S_L$ is expressed in terms of the fully dressed
Nambu Green's function $\hat{G}^{+,-}$, see Refs.~\cite{Caroli1971,Caroli1972,Cuevas1}:
\begin{equation}
  \label{eq:IL}
  I_L=\mbox{Nambu trace}\left\{\hat{\tau}_3\left[\hat{\Sigma}_{\alpha,a}
    \hat{G}^{+,-}_{a,\alpha} - \hat{\Sigma}_{a,\alpha}\hat{G}^{+,-}_{\alpha,a}
    \right]\right\}
  .
\end{equation}
Assuming a large superconducting gap and a continuous 2D metal density
of states leads to a finite value for the corresponding Keldysh
Green's function, which leads to the following expression of
$\hat{G}^{+,-}_{a,\alpha}$:
\begin{eqnarray}
  \hat{G}^{+,-}_{a,\alpha}&=&
  \hat{g}^{R}_{a,a} \hat{\Sigma}_{a,\alpha} \hat{g}^{+,-}_{\alpha,\beta}
  \hat{\Sigma}_{\beta,b} g^A_{b,c} \hat{\Sigma}_{c,\gamma}
  \hat{g}^A_{\gamma,\delta} \hat{\Sigma}_{\delta,d} \hat{g}^A_{d,d}
  \hat{\Sigma}_{d,\delta} \hat{g}^A_{\delta,\gamma} \hat{\Sigma}_{\gamma,c}
  \hat{g}^A_{c,b}\hat{\Sigma}_{b,\beta} \hat{g}^A_{\beta,\alpha}\\
  &+&
  \hat{g}^{R}_{a,a} \hat{\Sigma}_{a,\alpha} \hat{g}^{R}_{\alpha,\beta}
  \hat{\Sigma}_{\beta,b} g^R_{b,c} \hat{\Sigma}_{c,\gamma}
  \hat{g}^{+,-}_{\gamma,\delta} \hat{\Sigma}_{\delta,d} \hat{g}^A_{d,d}
  \hat{\Sigma}_{d,\delta} \hat{g}^A_{\delta,\gamma} \hat{\Sigma}_{\gamma,c}
  \hat{g}^A_{c,b}\hat{\Sigma}_{b,\beta} \hat{g}^A_{\beta,\alpha}\\
  &+&
  \hat{g}^{R}_{a,a} \hat{\Sigma}_{a,\alpha} \hat{g}^{R}_{\alpha,\beta}
  \hat{\Sigma}_{\beta,b} g^R_{b,c} \hat{\Sigma}_{c,\gamma}
  \hat{g}^R_{\gamma,\delta} \hat{\Sigma}_{\delta,d} \hat{g}^R_{d,d}
  \hat{\Sigma}_{d,\delta} \hat{g}^{+,-}_{\delta,\gamma} \hat{\Sigma}_{\gamma,c}
  \hat{g}^A_{c,b}\hat{\Sigma}_{b,\beta} \hat{g}^A_{\beta,\alpha}\\
  &+&
  \hat{g}^{R}_{a,a} \hat{\Sigma}_{a,\alpha} \hat{g}^{R}_{\alpha,\beta}
  \hat{\Sigma}_{\beta,b} g^R_{b,c} \hat{\Sigma}_{c,\gamma}
  \hat{g}^R_{\gamma,\delta} \hat{\Sigma}_{\delta,d} \hat{g}^R_{d,d}
  \hat{\Sigma}_{d,\delta} \hat{g}^R_{\delta,\gamma} \hat{\Sigma}_{\gamma,c}
  \hat{g}^R_{c,b}\hat{\Sigma}_{b,\beta} \hat{g}^{+,-}_{\beta,\alpha}
  .
\end{eqnarray}
Given the assumption of a large gap and the
forms~(\ref{eq:g2D-1})-(\ref{eq:g2D-2}) of the 2D metal Green's
functions, we obtain the following identities:
\begin{eqnarray}
  \label{eq:identity1}
  0&=&
 \hat{g}^{+,-}_{\alpha,\beta}
  \hat{\Sigma}_{\beta,b} g^A_{b,c} \hat{\Sigma}_{c,\gamma}
  \hat{g}^A_{\gamma,\delta}
  +
  \hat{g}^{R}_{\alpha,\beta}
  \hat{\Sigma}_{\beta,b} g^R_{b,c} \hat{\Sigma}_{c,\gamma}
  \hat{g}^{+,-}_{\gamma,\delta} \\
  0&=&
\hat{g}^{+,-}_{\delta,\gamma} \hat{\Sigma}_{\gamma,c}
  \hat{g}^A_{c,b}\hat{\Sigma}_{b,\beta} \hat{g}^A_{\beta,\alpha}
  +
 \hat{g}^{R}_{\delta,\gamma} \hat{\Sigma}_{\gamma,c}
  \hat{g}^R_{c,b}\hat{\Sigma}_{b,\beta} \hat{g}^{+,-}_{\beta,\alpha}
  ,
  \label{eq:identity2}
\end{eqnarray}
from what we deduce that the corresponding current is vanishingly
small: $I_L\equiv I_a=0$ because $\hat{G}^{+,-}_{a,\alpha} = 0$ and
$\hat{G}^{+,-}_{\alpha,a} = 0$ in Eq.~(\ref{eq:IL}). Identities
similar to Eqs.~(\ref{eq:identity1})-(\ref{eq:identity2}) are also
operational in elastic cotunneling (EC) at a normal
metal-superconductor-normal metal double interface, evaluated at the
lowest order in the tunneling amplitudes. In this case, it is known
\cite{Falci,Melin-Feinberg} that, in linear response, the current is
proportional to the difference between the right and left
electrochemical potentials, which implies that the EC current is
vanishingly small if both electrochemical potentials are identical, as
it is the case in the processes of scattering Andreev tubes on a
single interface.

We conclude that scattering Andreev tubes on the boundaries is not
effective in all-superconducting devices. Consequently, the Andreev
tubes are parallel to the $x$-axis in Andreev interferometers, which
is compatible with the experimental robustness of the coherent ABS
oscillations reported in Ref.~\cite{NonequilibriumABS2025}, and, in the
present paper, with the robustness of the quartet ABS oscillations as
the voltage $V_t$ is increased on the superconducting tunneling probe.

Conversely, the above argument can easily be generalized to scattering
Andreev tubes at a 2D normal metal-3D metal interface, and a
nonvanishingly small current is found in this case. The Andreev tube
then scatter at the normal contact interface, in a way that is
qualitatively similar to scattering on nonmagnetic impurities. The
resulting destructive interference between tubes of different lengths
is compatible with the absence of robustness of the coherent ABS
oscillations as the voltage is increased on the normal contact of a
$SSN$ Andreev interferometer, see Ref.~\cite{NonequilibriumABS2025}.

\end{document}